\documentclass[twocolumn, aps,prd,preprintnumbers,superscriptaddress,showpacs,nofootinbib]{revtex4-1}
\usepackage{amsmath,amsthm,amssymb,mathptmx}
\usepackage[dvipdfmx]{graphicx}
\usepackage{color}
\usepackage{float}
\usepackage{booktabs}
\usepackage{bm}

\allowdisplaybreaks[1]

\newcommand{\fnl}{f_{\rm NL}}
\newcommand{\gnl}{g_{\rm NL}}
\newcommand{\tnl}{\tau_{\rm NL}}
\newcommand{\png}{the primordial non-Gaussianity}

\newcommand{\pab}{the power spectra and the bispectra}

\newcommand{\angpab}{the angular-power spectra and bispectra}

\begin{document}
\title{Constraining higher-order parameters for primordial non-Gaussianities from power spectra and bispectra of imaging survey} 

\author{Ichihiko~Hashimoto}
\affiliation{Yukawa Institute for Theoretical Physics, Kyoto University, Kyoto 606-8502, Japan}

\author{Atsushi~Taruya}
\affiliation{Center for Gravitational Physics, Yukawa Institute for Theoretical Physics, Kyoto University, Kyoto 606-8502, Japan}
\affiliation{Kavli Institute for the Physics and Mathematics of the Universe (WPI), Todai institute
for Advanced Study, University of Tokyo, Kashiwa, Chiba 277-8568, Japan}

\author{Takahiko~Matsubara}
\affiliation{Department of Physics, Nagoya University, Chikusa, Nagoya, 464-8602, Japan}
\affiliation{Kobayashi-Maskawa Institute for the Origin of Particles and the Universe, Nagoya University, Chikusa, Nagoya, 464-8602, Japan}

\author{Toshiya~Namikawa}
\affiliation{Department of Physics, Stanford University, Stanford, CA 94305, USA}
\affiliation{Kavli Institute for Particle Astrophysics and Cosmology,
SLAC National Accelerator Laboratory, Menlo Park, CA 94025, USA}

\author{Shuichiro~Yokoyama}
\affiliation{Department of Physics,
Rikkyo University, 3-34-1 Nishi-Ikebukuro, Toshima, Tokyo 171-8501, Japan}

\date{\today}

\begin{abstract}
We investigate the statistical power of higher-order statistics and cross-correlation statistics to constrain the primordial non-Gaussianity from the imaging surveys. In particular, we consider the local-type primordial non-Gaussianity and discuss how well one can tightly constrain the higher-order non-Gaussian parameters ($g_{\rm NL}$ and $\tau_{\rm NL}$) as well as the leading order parameter $f_{\rm NL}$ from the halo/galaxy clustering and weak gravitational lensing measurements. Making use of a strong scale-dependent behavior in the galaxy/halo clustering, Fisher matrix analysis reveals that the bispectra can break the degeneracy between non-Gaussian parameters ($f_{\rm NL}$, $g_{\rm NL}$ and $\tau_{\rm NL}$) and this will give simultaneous constraints on those three parameters. The combination of cross-correlation statistics further improves the constraints by factor of $2$. As a result, upcoming imaging surveys like the Large Synoptic Survey Telescope have the potential to improve the constraints on the primordial non-Gaussianity much tighter than those obtained from the CMB measurement by Planck, giving us an opportunity to test the single-sourced consistency relation, $\tau_{\rm NL}\ge(36/25)f_{\rm NL}^2$.
\end{abstract}

\preprint{RUP-15-30}
\preprint{YITP-15-121}

\maketitle


\section{Introduction}
\label{sec:intro}

The observations of cosmic microwave background anisotropies 
give us many hints on the very early epoch of the Universe beyond the last 
scattering surface. So far, the measurements are pretty consistent with 
the inflationary scenario that the Universe had undergone an accelerated 
cosmic expansion during a short period of time, when the curvature 
perturbations as a seed of large-scale structure had been quantum-mechanically generated. With the increased precision data by Planck, many 
models of cosmic inflation are severely constrained or ruled out.  

Among various clues to test or clarify the early cosmic inflation, 
(non-)detection of primordial non-Gaussianity is a key to understand 
the generation mechanism of primordial density fluctuations.  
A solid theoretical prediction is that the 
single-field slow-roll inflation produces nearly 
Gaussian fluctuations, but multi-field inflations or nonlinear super-horizon 
dynamics can lead to a large non-Gaussianity \cite{2010JCAP...12..030S}. 
The measurement by Planck satellite currently gives non-detection of primordial 
non-Gassianity, and has derived the strong constraint, 
$f_{\rm NL}=0.71\pm 10.2$ at $2\sigma$ level \cite{2013arXiv1303.5084P}, where $f_{\rm NL}$ characterizes the amplitude of the bispectrum of the primordial 
curvature-fluctuations for a specific non-Gaussianity called local-type. 
Note that the local-type non-Gaussianity is known to arise from the non-linear dynamics of the primordial curvature perturbations on super-horizon scales. In general,  
it is characterized not only by $f_{\rm NL}$, but also the higher-order parameters such as $g_{\rm NL}$ and $\tau_{\rm NL}$, which are related to the shape and amplitude of the trispectrum of the primordial curvature fluctuations. Indeed, these higher-order parameters are helpful to further discriminate between many models of inflation. For instance, an inequality between $f_{\rm NL}$ and $\tau_{\rm NL}$ called Suyama-Yamaguchi inequality, $\tau_{\rm NL}\ge (36/25)f_{\rm NL}^2$ \cite{2008PhRvD..77b3505S} tells us that the equality $\tau_{\rm NL}= (36/25)f_{\rm NL}^2$ holds only if the single scalar field is the source of the primordial curvature fluctuation. Thus, the observational confirmation of this inequality immediately implies that the source of the primordial curvature fluctuations is multi-field. However, the constraints on the higher-order parameters $g_{\rm NL}$ and $\tau_{\rm NL}$ are still weak; $g_{\rm NL}=(-1.2\pm 2.8)\times 10^5$, $\tau_{\rm NL}<2800$ at $2\sigma$ level from the Planck observation \cite{2013arXiv1303.5084P}.

Alternative powerful probe may be the large-scale structure observations via 
the spectroscopic and/or photometric measurements. 
In the presence of local-type non-Gaussianity,  
of particular interesting feature is a strong enhancement of 
the halo/galaxy clustering amplitude on large scales, as has been 
recently revealed by numerical and theoretical studies (e.g., \cite{2008PhRvD..77l3514D,2008PhRvD..78l3507A}). Refs.~\cite{2011PhRvD..83l3514N,2012PhRvD..85d3518T} showed that the power spectrum measurement of the halo/galaxy clustering in future surveys like Large Synoptic Survey Telescope (LSST) can give a tighter constraint than that from the CMB measurement. However, non-zero values of $g_{\rm NL}$ and $\tau_{\rm NL}$ similarly lead to a scale-dependent enhancement of the clustering amplitude, and simultaneous constraints on $f_{\rm NL}$, $g_{\rm NL}$ and $\tau_{\rm NL}$ suffer from a strong parameter degeneracy. Hence, in order to avoid a substantial degradation of the constraining power on non-Gaussianity, other statistical information needs to be combined. One powerful approach may be to make use of the multi-tracer method (e.g., \cite{Seljak:2008xr,2011PhRvD..84h3509H}), with which we can take advantage of 
the different mass- and redshift-dependence of the non-Gaussian induced clustering \cite{Yamauchi:2015mja}.

In this paper, as another but solid approach,  
we examine the statistical power of combining the auto-/cross-power spectra and bispectra of halo/galaxy clustering and weak-lensing data, obtained from on-going/upcoming imaging surveys. 
The bispectrum of the halo/galaxy clustering is a direct probe of non-Gaussian feature, and the scale-dependent enhancement by local-type non-Gaussianity is known to appear in the bispectrum amplitude \cite{2014PhRvD..89d3524Y,2009ApJ...703.1230J,2010JCAP...07..002N,2011JCAP...04..006B}. Hence, the combination of power spectrum and bispectrum is expected to give a tighter constraint on the primordial non-Gaussianity \cite{2004PhRvD..69j3513S,2007PhRvD..76h3004S,2012MNRAS.425.2903S}. However, most of previous works focuses on the constraint on $f_{\rm NL}$ only. Further, compared to the auto-correlation statistics, the cross correlations between halo/galaxy clustering and weak-lensing fields have a different dependence on the primordial non-Gaussianity. We thus expect that the 
combination of bispectrum and cross correlation statistics can break parameter degeneracy, and it enables us to give tight constraints on $f_{\rm NL}$, $g_{\rm NL}$ and $\tau_{\rm NL}$ simultaneously.

In order to estimate the forecast constraints, we need a robust theoretical prediction for power spectra and bispectra, including not only primordial non-Gaussianity effect but also non-linear gravitational clustering which gives another source of non-Gaussianity. Here, we employ integrated-Perturbation Theory (iPT) to compute power spectra and bispectra \cite{2011PhRvD..83h3518M,2012PhRvD..86f3518M}. The iPT is the resummed PT formalism based on the multi-point propagator expansion \cite{2008PhRvD..78j3521B}, and it provides a framework to incorporate not only the nonlinear gravitational clustering but also the non-perturbative properties of halo/galaxy clustering bias into the theoretical calculation. With this iPT formalism, the calculation of higher-order statistics is straightforward at weakly non-linear regime. We show that the combination of the auto-/cross-power spectra and bispectra can break the degeneracy between the constraints on $f_{\rm NL}$, $g_{\rm NL}$ and $\tau_{\rm NL}$. As a result, upcoming surveys like LSST have the potential to improve the marginalized constraints on the primordial non-Gaussianity much tighter than those obtained from the Planck observation, giving us an opportunity to test Suyama-Yamaguchi inequality, $\tau_{\rm NL}\ge (36/25)f_{\rm NL}^2$ \cite{2008PhRvD..77b3505S}.

This paper is organized as follows. In Sec.\ref{sec:powerandbi} we briefly review the iPT and present the formulas for the three dimensional auto-/cross-power spectra and bispectra of halo/galaxy and matter distribution in the presence of \png. Then, in Sec.\ref{sec:obs}, we calculate the angular-power spectra and bispectra, which are observables of the photometric/imaging galaxy surveys. We then evaluate their covariance based on the formula of previous section. In Sec.\ref{sec:SN}, we estimate signal-to-noise ratio based on three representative future-surveys, Subaru Hyper Suprime-Cam (HSC), Dark Energy Survey (DES) and LSST. In Sec.\ref{sec:fifi}, based on the Fisher matrix formalism, we quantitatively estimate the impact of the bispectra and weak-gravitational lensing effect on the detection of primordial non-Gaussianity. Finally, Sec.\ref{sec:discuss} is devoted to the summary and discussion.

Throughout the paper, primordial power spectrum is computed with the  cosmological Boltzmann code, CAMB\cite{2000ApJ...538..473L}, with the cosmological parameters assuming a flat Lambda-CDM model, determined by WMAP9 results \cite{2013ApJS..208...19H}; $\Omega_{\rm b}=0.046$, $\Omega_{\rm c}=0.23$, $\Omega_{\Lambda 0}=0.72$, $n_{\rm s}=0.97$, $A_{\rm s}=2.4\times10^{-9}$ and $w=-1$, for the density parameters of baryon and cold dark matter, dark energy density, scalar spectral index, scalar amplitude at $k=0.002 [{\rm Mpc^{-1}}]$ and dark-energy equation-of-state parameter, respectively.

\section{Power spectrum and Bispectrum with primordial non-Gaussianity\label{sec:powerandbi}}

In this section, employing the integrated perturbation theory (iPT) proposed by 
Ref.~\cite{2011PhRvD..83h3518M}, we present the formulas for the auto-/cross-power spectra and bispectra of halo/galaxy and matter distribution in the presence of \png. In iPT, multipoint propagators are the building block of perturbative expansion. Given the initial condition, the evolved results of power spectra and bispectra at given redshift are systematically constructed with multi-point propagators. In particular, based on a halo bias prescription, iPT treatment enables us to incorporate the scale-dependent nature of halo bias into the prediction in a non-perturbative manner. After briefly reviewing the local-type non-Gaussianity as a representative model of primordial non-Gaussianity, in Sec.\ref{sec:local}, we give the perturbative expressions for \pab, in Sec.\ref{sec:ipt}. Then, Sec.\ref{sec:multi} presents the formula for multi-point propagators relevant at large scales of our interest.

\subsection{Local-type primordial non-Gaussianity\label{sec:local}}

According to the inflationary scenario,  the primordial curvature perturbations $\Phi$ as the seed of large-scale structure are quantum mechanically generated. 
Among various inflation models proposed so far, the single-field slow-roll inflation generically predicts the Gaussian fluctuations, and thus the statistical properties of primordial perturbations are solely characterized by power spectrum:
\begin{align}
\langle\Phi(\bm{k})\Phi(\bm{k}')\rangle&=(2\pi)^3\delta_{\rm D}^{(3)}(\bm{k}+\bm{k}')P_{\Phi}(k),
\end{align} 
where $\bm{k}$ and $\delta_{\rm D}^{(3)}$ are respectively three-dimensional wave vector and the three-dimensional Dirac's delta function. The bracket $\langle~\rangle$ means ensemble average.  

On the other hand, 
multi-field or non-slow-roll inflation models are known to produce non-Gaussianities for primordial perturbations \cite{2010CQGra..27l4002W,2010CQGra..27l4001K}, and their non-Gaussian nature is characterized by the non-vanishing higher-order spectra of primordial curvature perturbations such as bispectrum and trispectrum: 
\begin{align}
&\langle\Phi(\bm{k}_1)\Phi(\bm{k}_2)\Phi(\bm{k}_3)\rangle_{\rm c}
\nonumber\\
&\qquad ~=(2\pi)^3\,\delta_{\rm D}^{(3)}(\bm{k}_1+\bm{k}_2+\bm{k}_3)\,B_{\Phi}(\bm{k}_1,\bm{k}_2,\bm{k}_3),\\
&\langle\Phi(\bm{k}_1)\Phi(\bm{k}_2)\Phi(\bm{k}_3)\Phi(\bm{k}_4)\rangle_{\rm c}
\nonumber\\
&\qquad ~=(2\pi)^3\,\delta_{\rm D}^{(3)}(\bm{k}_1+\bm{k}_2+\bm{k}_3+\bm{k}_4)\,T_{\Phi}(\bm{k}_1,\bm{k}_2,\bm{k}_3,\bm{k}_4),
\end{align} 
where the subscript, ${\rm c}$, denotes the connected part of the correlation functions. 

In this paper, among several types of non-Gaussianities known so far, we are particularly interested in the local-type non-Gaussianity, originated from non-linear dynamics of the primordial curvature perturbations on super-horizon scales. In this model, the primordial curvature perturbation $\Phi$ is described by the Taylor-series expansion of Gaussian field $\Phi_{\rm G}$ as
\begin{align}
\Phi(\bm{x})=\Phi_{\rm G}(\bm{x})+\fnl\left\{\Phi_{\rm G}(\bm{x})^2-\langle\Phi_{\rm G}(\bm{x})^2\rangle\right\}+\gnl\Phi_{\rm G}(\bm{x})^3+\cdots.\label{phhhp}
\end{align}
With this description, the higher-order spectra are expressed as
\begin{align}
&B_{\rm \Phi}\left(k_1,k_2,k_3\right)\simeq 2\fnl\left\{P_{\Phi}\left(k_1\right)P_\Phi\left(k_2\right)
\right.
\nonumber\\
& \qquad\qquad\qquad\qquad\quad\quad
\left.
+\,\,\mbox{2\,perms}\,\,
(k_1\leftrightarrow k_2\leftrightarrow k_3)\right\},
\label{eq:localbi}
\\
&T_{\rm \Phi}\left(k_1,k_2,k_3,k_4\right)\simeq 6\gnl\left\{P_{\Phi}\left(k_1\right)P_\Phi\left(k_2\right)P_{\Phi}\left(k_3\right)+\,\mbox{3\,perms}\,\right\}\nonumber\\
&\quad+\frac{25}{9}\tnl\left\{P_{\Phi}\left(k_1\right)P_\Phi\left(k_2\right)P_{\Phi}\left(|\bm{k}_1+\bm{k}_3|\right)
+11\,{\rm perms}\right\},
\label{eq:localtris}
\end{align}
where the parameters $\fnl,\gnl$ and $\tnl$ describe the strength of \png. These parameters are not independently given, but are known to be related each other according to the generation mechanisms of primordial non-Gaussianity. For the single-sourced curvature perturbations described by Eq.~(\ref{phhhp}), the parameter $\tnl$ is related to the leading-order parameter $\fnl$ through $\tnl=(36/25)\fnl^2$. For multi-sourced curvature perturbation, on the other hand, the inequality, $\tnl > (36/25)\fnl^2$, generally holds \cite{2008PhRvD..77b3505S}. These relations are called Suyama-Yamaguchi inequality. The observational probe of this inequality thus provides an important clue for the scalar-field dynamics during the inflation.

Provided the statistical nature of primordial curvature perturbations, 
the linear density field $\delta$ is determined through 
\begin{align}
\delta_{\rm L}(\bm{k},z)&=M(k,z)\Phi(\bm{k},z_*),\label{dell}
\end{align}
with the function $M(k,z)$ given by 
\begin{align}
M(k,z)&=\frac{2}{3}\frac{D(z)}{D(z_*)(1+z_*)}\frac{k^2T(k)}{H_0^2\Omega_{m0}},
\label{delldell}
\end{align}
where the function $T(k)$ and $D(z)$ are respectively the transfer function and the linear growth factor. The $H_0,\Omega_{m0}$ and $z_*$ are the Hubble parameter at present time, the matter density parameter, and the redshift at initial time in the matter-dominated era. Then, the power spectrum, bispectrum and trispectrum of the linear-density field are defined by
\begin{align}
&\langle\delta_{\rm L}(\bm{k}_1)\delta_{\rm L}(\bm{k}_2)\rangle=(2\pi)^3\delta_{\rm D}^{(3)}(\bm{k}_1+\bm{k}_2)\,P_{\rm L}(k_1),\\
&\langle \delta_{\rm L}(\bm{k}_1)\delta_{\rm L}(\bm{k}_2)\delta_{\rm L}(\bm{k}_3)\rangle
\nonumber\\
&\quad\quad=(2\pi)^3\delta_{\rm D}^{(3)}(\bm{k}_1+\bm{k}_2+\bm{k}_3)\,B_{\rm L}(k_1,k_2,k_3),\\
&\langle \delta_{\rm L}(\bm{k}_1)\delta_{\rm L}(\bm{k}_2)\delta_{\rm L}(\bm{k}_3)\delta_{\rm L}(\bm{k}_4)\rangle
\nonumber\\
&\quad\quad=(2\pi)^3\delta_{\rm D}^{(3)}(\bm{k}_1+\bm{k}_2+\bm{k}_3+\bm{k}_4)\,T_{\rm L}(k_1,k_2,k_3,k_4)
\end{align} 
From Eqs.~(\ref{dell}) and (\ref{delldell}), the higher-order spectra of density field are directly related to those of the primordial-curvature perturbations through: 
\begin{align}
&P_{\rm L}\left(k\right)=M\left(k\right)^2P_{\Phi}\left(k\right),\\
&B_{\rm L}\left(k_1,k_2,k_3\right)=M\left(k_1\right)M\left(k_2\right)M\left(k_3\right)B_{\Phi}\left(k_1,k_2,k_3\right),\label{eq:localbi2}\\
&T_{\rm L}\left(k_1,k_2,k_3,k_4\right)=M\left(k_1\right)M\left(k_2\right)M\left(k_3\right)M\left(k_4\right)T_{\Phi}\left(k_1,k_2,k_3,k_4\right).
\label{eq:localtris_2}
\end{align}

The linear density field is, however, indirectly related to the observable of large-scale structure, and we need to know a explicit relation between the linear density field and the observables especially probed with imaging surveys,   
taking a full account of the late-time gravitational evolution and the effect of galaxy/halo bias.

\subsection{Integrated Perturbation Theory (iPT) \label{sec:ipt}}

The large-scale structure can be a good probe of \png, and recent theoretical studies have revealed that the local-type non-Gaussianity can induce a scale-dependent enhancement of the clustering amplitude on large scales (e.g., \cite{2008PhRvD..77l3514D,2008PhRvD..78l3507A}). Therefore, in this paper, we focus on the clustering features of halo/galaxy density field obtained from the imaging surveys. Further, to enhance the statistical significance, we consider the weak gravitational lensing and cross-correlate it with halo/galaxy density field. To observationally probe \png~from large-scale structure, a proper account of the late-time gravitational evolution and the effect of halo/galaxy bias is essential. Here, we adopt integrated perturbation theory (iPT),  which enables us to simultaneously incorporate these two effects into the theoretical predictions. Let us first define the three-dimensional power spectra $\mathcal{P}_{\rm XY}$ and bispectra $\mathcal{B}_{\rm XYZ}$ 
of the observables: 
\begin{align}
&\langle \delta_{\rm X}(\bm{k}_1)\delta_{\rm Y}(\bm{k}_2)\rangle=
(2\pi)^3\delta_{\rm D}^{(3)}(\bm{k}_1+\bm{k}_2)\,\mathcal{P}_{\rm XY}(k),
\label{power3}\\
&\frac{1}{3}\left\{\langle \delta_{\rm X}(\bm{k}_1)\delta_{\rm Y}(\bm{k}_2)\delta_{\rm Z}(\bm{k}_3)\rangle+\mbox{2\,perms}\,(\bm{k}_1\leftrightarrow \bm{k}_2\leftrightarrow\bm{k}_3)\right\}
\nonumber\\
&\qquad\qquad=(2\pi)^3\delta_{\rm D}^{(3)}(\bm{k}_1+\bm{k}_2+\bm{k}_3)\,\mathcal{B}_{\rm XYZ}(k_1,k_2,k_3)
,\label{bis3}
\end{align}
where $\delta_{\rm X,Y,Z}$ is the three-dimensional density field. For the quantities relevant to the imaging surveys, we consider the two types of three-dimensional objects; one is the halo/galaxy density field $\delta_{\rm h}$, and the other is the matter fluctuation $\delta_{\rm m}$. In iPT,  the perturbative expansion of the statistical quantities such as power spectra and bispectra are constructed with multi-point propagators and the linear polyspectra 
(e.g., \cite{2011PhRvD..83h3518M,2012PhRvD..86f3518M}). Following Refs.~\cite{2011PhRvD..83h3518M,2012PhRvD..86f3518M}, we define the $(n+1)$-point propagators of the objects, $\Gamma_{\rm X}^{(n)}$ as 
\begin{align}
&\left\langle \frac{\delta^n\delta_{\rm X}\left(\bm{k}\right)}{\delta\delta_{\rm L}\left(\bm{k}_1\right)\delta\delta_{\rm L}\left(\bm{k}_2\right)\cdots\delta\delta_{\rm L}\left(\bm{k}_n\right)}\right\rangle
\nonumber\\
&\quad\quad=
\left(2\pi\right)^{3-3n}\delta_{\rm D}^{(3)}\left(\bm{k}-\bm{k}_{12\cdots n}\right)\Gamma_{\rm X}^{\left(n\right)}\left(\bm{k}_1,\bm{k}_2,\cdots,\bm{k}_n\right)
\label{gamma},
\end{align}
where $\delta_{\rm L}$ is the linear-density field. The multi-point propagators are fully non-perturbative quantities characterizing the non-linear gravitational evolution and halo/galaxy bias properties. With these propagators, power spectra and bispectra are generally expanded as
\begin{align}
&\mathcal{P}_{\rm XY}(k)=\mathcal{P}^{\rm tree}_{\rm XY}(k)+\mathcal{P}^{\rm 1-loop}_{\rm XY}(k)+\mathcal{P}^{\rm 2-loop}_{\rm XY}(k)+\cdots,
\label{eq:power3D}\\
&\mathcal{B}_{\rm XYZ}(k_1,k_2,k_3)=\mathcal{B}^{\rm tree}_{\rm XYZ}(k_1,k_2,k_3)+\mathcal{B}^{\rm 1-loop}_{\rm XYZ}(k_1,k_2,k_3)
\nonumber\\
&\qquad\qquad\qquad\qquad+\mathcal{B}^{\rm 2-loop}_{\rm XYZ}(k_1,k_2,k_3)+\cdots.\label{bis3D}
\end{align}
Here, the quantities $\mathcal{P}^{\rm n-loop}_{\rm XY}$ and $\mathcal{B}^{\rm n-loop}_{\rm XYZ}$ indicate the higher-order corrections which include $n$-loop integrals. Up to the one-loop order (i.e., next-to-leading order), the power spectra of the objects $X$ and $Y$ are systematically constructed from the contributions that do not vanish even with the Gaussian initial condition, $\mathcal{P}_{\rm grav}$, and the one originating from primordial bispectrum $\mathcal{P}_{\rm bis}$ \cite{2013PhRvD..87b3525Y}:
\begin{align}
\mathcal{P}^{\rm tree}_{\rm XY}=\mathcal{P}^{\rm tree}_{\rm grav}~,\qquad 
\mathcal{P}^{\rm 1-loop}_{\rm XY}=\mathcal{P}^{\rm 1-loop}_{\rm grav}+
\mathcal{P}^{\rm 1-loop}_{\rm bis}\label{Ptreee}
\end{align}
Each term at the right-hand side of this expressions is given by
\begin{widetext}
\begin{align}
&\mathcal{P}^{\rm tree}_{\rm grav}(k)=\Gamma_{\rm X}^{(1)}\Gamma_{\rm Y}^{(1)}P_{\rm L}(k),\\
&\mathcal{P}^{\rm 1-loop}_{\rm grav}(k)=\frac{1}{2}\int\frac{d^3 p}{(2\pi)^3}\Gamma_{\rm X}^{(2)}(\bm{p},\bm{k}-\bm{p})\Gamma_{\rm Y}^{(2)}(\bm{p},\bm{k}-\bm{p})P_{\rm L}(k)P_{\rm L}(|\bm{k}-\bm{p}|),\label{1-loopgravp}\\
&\mathcal{P}_{\rm bis}^{\rm 1-loop}(k)=\frac{1}{2}\int\frac{d^3p}{(2\pi)^3}\left\{\Gamma_{\rm X}^{(1)}(\bm{k})\Gamma_{\rm Y}^{(2)}(\bm{p},\bm{k}-\bm{p})+\Gamma_{\rm Y}^{(1)}(\bm{k})\Gamma_{\rm X}^{(2)}(\bm{p},\bm{k}-\bm{p})\right\}B_{\rm L}(\bm{k},-\bm{p},-\bm{k}+\bm{p}).\label{pppppppp}
\end{align}
\end{widetext}

Similarly, the bispectra at the one-loop order are constructed from the contributions of the gravity-induced non-Gaussian term $\mathcal{B}_{\rm grav}$, and the contributions from the primordial bispectrum and trispectrum, respectively denoted by $\mathcal{B}_{\rm bis}$ and $\mathcal{B}_{\rm tris}$ \cite{2014PhRvD..89d3524Y}:
\begin{align}
\mathcal{B}_{\rm XYZ}^{\rm tree}&=\mathcal{B}^{\rm tree}_{\rm grav}+\mathcal{B}^{\rm tree}_{\rm bis},
\label{eq:btree}\\
\mathcal{B}_{\rm XYZ}^{\rm 1-loop}&=\mathcal{B}^{\rm 1-loop,1}_{\rm grav}+
\mathcal{B}^{\rm 1-loop,2}_{\rm grav}+\mathcal{B}^{\rm 1-loop,1}_{\rm bis}
\nonumber\\
&+\mathcal{B}^{\rm 1-loop,2}_{\rm bis}+\mathcal{B}^{\rm 1-loop,3}_{\rm bis}+\mathcal{B}^{\rm 1-loop}_{\rm tris}.
\label{eq:b1loop}
\end{align}
The expressions for each contribution are explicitly given by
\begin{widetext}
\begin{align}
\mathcal{B}^{\rm tree}_{\rm grav}\left(\bm{k}_1,\bm{k}_2,\bm{k}_3\right)=&\frac{1}{3}\Bigl[\Bigl\{\Gamma^{\left(1\right)}_{\rm X}\left(\bm{k}_1\right)\Gamma^{\left(1\right)}_{\rm Y}\left(\bm{k}_2\right)\Gamma^{\left(2\right)}_{\rm Z}\left(-\bm{k}_1,-\bm{k}_2\right)P_{\rm L}\left(k_1\right)P_{\rm L}\left(k_2\right)\nonumber\\
&+2{\rm perms}({\rm X\leftrightarrow Y\leftrightarrow Z})\Bigr\}+2{\rm perms}(\bm{k}_1\leftrightarrow \bm{k}_2\leftrightarrow \bm{k}_3)\Bigr],\label{treegrav}\\
\mathcal{B}^{\rm tree}_{\rm bis}\left(\bm{k}_1,\bm{k}_2,\bm{k}_3\right)=&\Gamma^{\left(1\right)}_{\rm X}\left(\bm{k}_1\right)\Gamma^{\left(1\right)}_{\rm Y}\left(\bm{k}_2\right)\Gamma^{\left(1\right)}_{\rm Z}\left(\bm{k}_3\right)B_{\rm L}\left(\bm{k}_1,\bm{k}_2,\bm{k}_3\right),\label{treebis}\\
\mathcal{B}^{\rm 1-loop,1}_{\rm grav}\left(\bm{k}_1,\bm{k}_2,\bm{k}_3\right)=&\frac{1}{3}\Biggl[\int\frac{d^3p}{\left(2\pi\right)^3}\Gamma_{\rm X}^{\left(2\right)}\left(\bm{p},\bm{k}_1-\bm{p}\right)\Gamma^{\left(2\right)}_{\rm Y}\left(-\bm{p},\bm{k}_2+\bm{p}\right)\Gamma^{\left(2\right)}_{Z}\left(-\bm{k}_1+\bm{p},-\bm{k}_2-\bm{p}\right)\nonumber\\
&\times P_{\rm L}\left(p\right)P_{\rm L}\left(|\bm{k}_1-\bm{p}|\right)P_{\rm L}\left(|\bm{k}_2+\bm{p}|\right)+2{\rm perms}(\bm{k}_1\leftrightarrow \bm{k}_2\leftrightarrow \bm{k}_3)\Biggr],\label{loop1grav}\\
\mathcal{B}_{\rm grav}^{\rm 1-loop,2}\left(\bm{k}_1,\bm{k}_2,\bm{k}_3\right)=&\frac{1}{3}\Biggl[\Biggl\{\Gamma^{\left(1\right)}_{\rm X}\left(\bm{k}_1\right)P_{\rm L}\left(k_1\right)\int\frac{d^3p}{\left(2\pi\right)^3}\Gamma^{\left(2\right)}_{\rm Y}\left(\bm{p},\bm{k}_2-\bm{p}\right)\Gamma^{\left(3\right)}_{\rm Z}\left(-\bm{k}_1,-\bm{p},-\bm{k}_2+\bm{p}\right)\nonumber\\
&\times P_{\rm L}\left(p\right)P_{\rm L}\left(|\bm{k}_2-\bm{p}|\right)+2{\rm perms}({\rm X\leftrightarrow Y\leftrightarrow Z})\Biggr\}+5{\rm perms}(\bm{k}_1\leftrightarrow \bm{k}_2\leftrightarrow \bm{k}_3)\Biggr],\label{loop2grav}\\
\mathcal{B}_{\rm bis}^{\rm 1-loop,1}\left(\bm{k}_1,\bm{k}_2,\bm{k}_3\right)=&\frac{1}{3}\Biggl[\Biggl\{\Gamma^{\left(1\right)}_{\rm X}\left(\bm{k}_1\right)\Gamma^{\left(1\right)}_{\rm Y}\left(\bm{k}_2\right)\int\frac{d^3p}{\left(2\pi\right)^3}\Gamma^{\left(3\right)}_{\rm Z}\left(-\bm{k}_1,\bm{p},-\bm{k}_2-\bm{p}\right)P_{\rm L}\left(k_1\right)B_{\rm L}\left(\bm{k}_2,\bm{p},-\bm{k}_2-\bm{p}\right)\nonumber\\
&+2{\rm perms}({\rm X\leftrightarrow Y\leftrightarrow Z})\Biggr\}+5{\rm perms}(\bm{k}_1\leftrightarrow \bm{k}_2\leftrightarrow \bm{k}_3)\Biggr],\label{loop1bis}\\
\mathcal{B}_{\rm bis}^{\rm 1-loop,2}\left(\bm{k}_1,\bm{k}_2,\bm{k}_3\right)=&\frac{1}{3}\Biggl[\Biggl\{\Gamma^{\left(1\right)}_{\rm X}\left(\bm{k}_1\right)\Gamma^{\left(2\right)}_{\rm Y}\left(-\bm{k}_1,-\bm{k}_2\right)\int\frac{d^3p}{\left(2\pi\right)^3}\Gamma^{\left(2\right)}_{\rm Z}\left(\bm{p},\bm{k}_2-\bm{p}\right)P_{\rm L}\left(k_1\right)B_{\rm L}\left(-\bm{k}_2,\bm{p},\bm{k}_2-\bm{p}\right)\nonumber\\
&+2{\rm perms}({\rm X\leftrightarrow Y\leftrightarrow Z})\Biggr\}+5{\rm perms}(\bm{k}_1\leftrightarrow \bm{k}_2\leftrightarrow \bm{k}_3)\Biggr],\label{loop2bis}\\
\mathcal{B}_{\rm bis}^{\rm 1-loop,3}\left(\bm{k}_1,\bm{k}_2,\bm{k}_3\right)=&\frac{1}{3}\Biggl[\Biggl\{\Gamma^{\left(1\right)}_{\rm X}\left(\bm{k}_1\right)\int\frac{d^3p}{\left(2\pi\right)^3}\Gamma^{\left(2\right)}_{\rm Y}\left(\bm{p},\bm{k}_2-\bm{p}\right)\Gamma^{\left(2\right)}_{\rm Z}\left(-\bm{p},\bm{k}_3+\bm{p}\right)P_{\rm L}\left(p\right)B_{\rm L}\left(\bm{k}_1,\bm{k}_2+\bm{p},\bm{k}_3-\bm{p}\right)\nonumber\\
&+2{\rm perms}({\rm X\leftrightarrow Y\leftrightarrow Z})\Biggr\}+2{\rm perms}(\bm{k}_1\leftrightarrow \bm{k}_2\leftrightarrow \bm{k}_3)\Biggr],\label{loop3bis}\\
\mathcal{B}_{\rm tris}^{\rm 1-loop}\left(\bm{k}_1,\bm{k}_2,\bm{k}_3\right)=&\frac{1}{3}\Biggl[\Biggl\{\frac{1}{2}\Gamma^{\left(1\right)}_{\rm X}\left(\bm{k}_1\right)\Gamma^{\left(1\right)}_{\rm Y}\left(\bm{k}_2\right)\int\frac{d^3p}{\left(2\pi\right)^3}\Gamma^{\left(2\right)}_{\rm Z}\left(\bm{p},\bm{k}_3-\bm{p}\right)T_{\rm L}\left(\bm{k}_1,\bm{k}_2,\bm{p},\bm{k}_3-\bm{p}\right)\nonumber\\
&+2{\rm perms}({\rm X\leftrightarrow Y\leftrightarrow Z})\Biggr\}+2{\rm perms}(\bm{k}_1\leftrightarrow \bm{k}_2\leftrightarrow \bm{k}_3)\Biggr].\label{tris}
\end{align}
\end{widetext}

In many cases of perturbative calculations, higher-loop contributions are usually suppressed especially at large scales. It is, however, found by Ref.~\cite{2014PhRvD..89d3524Y} that local-type non-Gaussianity induces a strong scale-dependent enhancement on higher-loop contributions, and some of the two-loop contributions can eventually exceed the tree-level and one-loop contributions at very large scales \cite{2013PhRvD..87b3525Y,2014PhRvD..89d3524Y}. These non-negligible corrections are expressed as
\begin{widetext}
\begin{align}
\mathcal{P}_{\rm tris}^{\rm 2-loop}(k)=&\frac{1}{4}\int\frac{d^3p_1d^3p_2}{(2\pi)^6}\Gamma_{\rm X}^{(2)}(\bm{p}_1,\bm{k}-\bm{p}_1)\Gamma_{\rm Y}^{(2)}(-\bm{p}_2,-\bm{k}+\bm{p}_2)T_{\rm L}(\bm{p}_1,\bm{k}-\bm{p}_1,-\bm{p}_2,-\bm{k}+\bm{p}_2)\nonumber\\
+&\frac{1}{6}\int\frac{d^3p_1d^3p_2}{(2\pi)^6}\left\{\Gamma_{\rm X}^{(1)}(\bm{k})\Gamma_{\rm Y}^{(3)}(\bm{p}_1,\bm{p}_2,\bm{k}-\bm{p}_1-\bm{p}_2)+\Gamma_{\rm Y}^{(1)}(\bm{k})\Gamma_{\rm X}^{(3)}(\bm{p}_1,\bm{p}_2,\bm{k}-\bm{p}_1-\bm{p}_2)\right\}\nonumber\\
&\times T_{\rm L}(\bm{k},-\bm{p}_1,-\bm{p}_2,-\bm{k}+\bm{p}_1+\bm{p}_2)
\label{2-looppower}
\end{align}
for the power spectrum, and 
\begin{align}
\mathcal{B}_{\rm tris}^{\rm 2-loop}\left(\bm{k}_1,\bm{k}_2,\bm{k}_3\right)=&\frac{1}{4}\Bigl[\Bigl\{\Gamma_{\rm X}^{(1)}(\bm{k}_1)P_{\rm L}(k_1)\int\frac{d^3p_1d^3p_2}{(2\pi)^6}\Gamma_{\rm Y}^{(2)}(\bm{p}_1,\bm{k}_2-\bm{p}_1)\Gamma_{\rm Z}^{(3)}(\bm{k}_1,\bm{p}_2,-\bm{k}_2-\bm{p}_2)\nonumber\\
&\times T_{\rm L}(\bm{p}_1,\bm{k}_2-\bm{p}_1,\bm{p}_2,-\bm{k}_2-\bm{p}_2)+2{\rm perms}({\rm X\leftrightarrow Y\leftrightarrow Z})\Bigr\}\nonumber\\
&+5{\rm perms}(\bm{k}_1\leftrightarrow \bm{k}_2\leftrightarrow \bm{k}_3)\Bigr]
\label{2-loopbi}
\end{align}
for the bispectrum. 
\end{widetext}

The other two-loop contributions are shown to be negligible at observable scales \cite{2013PhRvD..87b3525Y,2014PhRvD..89d3524Y}. The contributions given above come from the most dominant two-loop {\it un-decomposable} diagrams. In this paper, we take account of these contributions in computing \pab~of biased objects.

\subsection{The multi-point propagator at large scales\label{sec:multi}}

The multi-point propagator is defined as fully non-perturbative quantity, and it is generally difficult to evaluate it rigorously. At large scales ($k\rightarrow 0$), however, non-linearity of gravitational evolution is weak, and the perturbative treatment of multipoint propagators works well. According to Ref.~\cite{2014PhRvD..89d3524Y}, we have 
\begin{align}
\Gamma^{(1)}_{\rm X}(\bm{k})&\simeq 1+c^{\rm L}_1(k),\label{gamma1}\\
\Gamma^{(2)}_{\rm X}(\bm{k}_1,\bm{k}_2)&\simeq F_2(\bm{k}_1,\bm{k}_2)+\left(1+\frac{\bm{k}_1\cdot\bm{k}_2}{k_2^2}\right)c_1^{\rm L}(\bm{k}_1)
\nonumber\\
&+\left(1+\frac{\bm{k}_1\cdot\bm{k}_2}{k_1^2}\right)c_1^{\rm L}(\bm{k}_2)+c_2^{\rm L}(\bm{k}_1,\bm{k}_2). \label{gamma2}
\end{align}
Here, $F_2$ and $c_n^{\rm L}$ are respectively the second-order kernel of standard perturbation theory and a renormalized-bias function defined in Lagrangian space. Note that $c_n^{\rm L}=0$ in the case of mass fluctuation (i.e., ${\rm X}={\rm m}$). The explicit expressions for these quantities are given by 
\begin{align}
F_2(\bm{k}_1,\bm{k}_2)&=\frac{10}{7}+\left(\frac{k_2}{k_1}+\frac{k_1}{k_2}\right)\frac{\bm{k}_1\cdot\bm{k}_2}{k_1k_2}+\frac{4}{7}\left(\frac{\bm{k}_1\cdot\bm{k}_2}{k_1k_2}\right)^2,\\
c_n^{\rm L}(\bm{k}_1,\bm{k}_2,\cdots,\bm{k}_n)&=(2\pi)^{3n}\int\frac{d^3k'}{(2\pi)^3}
\nonumber\\
&\quad\times
\left\langle \frac{\delta^n\delta_{\rm X}^{\rm L}(\bm{k}')}{\delta\delta_{\rm L}(\bm{k}_1)\delta\delta_{\rm L}(\bm{k}_2)\cdots\delta\delta_{\rm L}(\bm{k}_n)}\right\rangle,
\label{cnl}
\end{align}
where the $\delta_{\rm X}^{\rm L}$ is the number-density fluctuation of the biased object ${\rm X}$ in Lagrangian space. When the wave number of our interest is sufficiently smaller than that of the integration variable ($k\ll p$), the multi-point propagators are approximately described by \cite{2011PhRvD..83h3518M}
\begin{align}
\Gamma^{(2)}_{\rm X}(\bm{p},\bm{k}-\bm{p})&\simeq\Gamma^{(2)}_{\rm X}(\bm{p},-\bm{p})\simeq c_2^{\rm L}(\bm{p},-\bm{p}),
\label{gamma2_kp}\\
\Gamma^{(3)}_{\rm X}(-\bm{k}_1,-\bm{p},-\bm{k}_2+\bm{p})&\simeq \frac{\bm{k}_1\cdot(\bm{k}_1+\bm{k}_2)}{k_1^2}c_2^{\rm L}(-\bm{p},\bm{p})
\nonumber\\
&+c_3^{\rm L}(-\bm{k}_1,-\bm{p},-\bm{k}_2+\bm{p}).\label{gamma3}
\end{align}
The approximations given in 
Eqs.~(\ref{gamma1}), (\ref{gamma2}), (\ref{gamma2_kp}) and (\ref{gamma3}) 
 are called large-scale limit. In Appendix.\ref{sec:largelim}, we present expressions of \pab~in the large-scale limit, assuming the local-type primordial non-Gaussianity.

For a quantitative calculation, expressions for the renormalized bias function is further needed. Here, we adopt the halo-bias prescription proposed by \cite{2011PhRvD..83h3518M}. Then, the renormalized bias function for halos with mass $M$ is computed from the halo mass function:
\begin{align}
&c_n^{\rm L}(\bm{k}_1,\cdots,\bm{k}_n)=\frac{A_n(M)}{\delta_{\rm c}^n}W(k_1,M)\cdots W(k_n,M)
\nonumber\\
&\quad+\frac{A_{n-1}(M)\sigma^n_M}{\delta_{\rm c}^n}
\frac{d}{d\ln{\sigma_M}}\left[\frac{W(k_1,M)\cdots W(k_n,M)}{\sigma^n_M}\right]
\label{cnl-wind}
\end{align}
with the coefficient $A_n$ given by 
\begin{align}
& A_n(M)\equiv\sum^n_{j=0}\frac{n!}{j!}\delta_{\rm c}^j\,
(-\sigma_M)^{-j}f_{\rm MF}^{-1}(\nu)\frac{d^jf_{\rm MF}(\nu)}{d\nu^j}. 
\end{align}
Here the quantity $\delta_{\rm c}(\simeq 1.68)$ is the so-called critical density of the spherical collapse model, $W(k,M)$ is top-hat window function over mass scale $R=(3M/4\pi\rho_{\rm m})^{1/3}$, and $\rho_{\rm m}$ is the matter density. The quantity $\sigma_{\rm M}$ is the dispersion of smoothed matter density field over mass scales:
\begin{align}
\sigma_{\rm M}^2=\int\frac{k^2dk}{2\pi^2}W^2(k,M)P_{\rm L}(k).
\end{align}
The function $f_{\rm MF}(\nu)$ is defined through the halo mass function $n(M,z)$:
\begin{align}
\nu f_{\rm MF}(\nu)&\equiv M^2\frac{n(M,z)}{\bar{\rho}}\frac{d\log{M}}{d\log{\nu}}\label{MF},
\end{align}
where $\nu=\delta_{\rm c}/\sigma_M$. Throughout the paper, we adopt the Sheth-Tormen fitting formula for the halo mass function $n(M,z)$ \cite{2001MNRAS.323....1S}:
\begin{align}
f_{\rm ST}(\nu)&=A(p)\sqrt{\frac{2}{\pi}}[1+(q\nu^2)^{-p}]\sqrt{q}\nu e^{-q\nu^2/2},
\label{st}
\end{align}
with $p=0.3$, $q=0.707$ and $A(p)=[1+\pi^{-1/2}2^{-p}\Gamma(1/2-p)]^{-1}$. $\Gamma(x)$ is the Gamma function. 

\section{Observables of imaging survey and error covariance\label{sec:obs}}

To elucidate the statistical power of bispectrum and cross correlation between halo and weak-lensing fields obtained from imaging surveys, we shall consider three representative surveys: HSC, DES and LSST. The statistical quantities observed with these surveys are the angular-power spectra and bispectra projected on the celestial sphere. In Sec.\ref{sec:angpabis}, we derive the formula for angular-power spectra and bispectra, which are related to the three-dimensional counterpart given in previous section. Then, in Sec.\ref{sec:covmat}, we present the error covariance of these quantities, which will be later used to estimate the signal-to-noise ratio and statistical uncertainties of the non-Gaussian parameters in subsequent section.

\subsection{Imaging surveys\label{threesurbey}}

As representative ongoing/upcoming imaging surveys, we consider HSC for deep, and DES for wide, and LSST for an idealistically deep and wide surveys. These surveys are characterized by the survey area $f_{\rm sky}\equiv\Omega_{\rm s}/4\pi$, the mean source redshift $z_{\rm m}$, and the mean number density of source galaxies par unit area $\bar{n}_{\rm s}$, which are summarized in TABLE \ref{ngsmfsky}.

\begin{table}[tb]
\begin{ruledtabular}
  \begin{tabular}{lllc}
     & $f_{\rm sky}$ & $z_{\rm m}$ &\quad $\bar{n}_{\rm s}\quad [{\rm arcmin}^{-2}]$ \\ \hline
    HSC~\cite{HSCrev} &  $0.0375~(1,500\,{\rm deg}^2)$ &  $1.0$&  $35$\\ \hline
    DES~\cite{2005astro.ph.10346T} &  $0.125~(5,000\,{\rm deg}^2)$ &  $0.5$&  $12$\\ \hline
    LSST~\cite{2009arXiv0912.0201L} &  $0.5~(20,000\,{\rm deg}^2)$ &    $1.5$&  $100$\\ 
  \end{tabular}
  \caption{Specification of survey parameters for three representative imaging surveys; sky coverage $f_{\rm sky}$, mean source redshift $z_{\rm m}$ and mean number density of source galaxies $\bar{n}_{\rm s}$. }
  \label{ngsmfsky}
  \end{ruledtabular}
\end{table}

To calculate expected signals for weak-gravitational lensing, we need the redshift distribution of source galaxies. Here, we adopt the following functional form for redshift distribution (e.g., \cite{2011PhRvD..83l3514N}): 
\begin{align}
 n_{\rm s}(z)dz=\bar{n}_{\rm s}\frac{3z^2}{2(0.64z_{\rm m})^3}\exp{\left[-\left(\frac{z}{0.64z_{\rm m}}\right)^{3/2}\right]}dz.\label{keiken}
\end{align}
Note that this will later appear in the weight function for weak-lensing observables, $W_{\kappa}$ [Eq.~(\ref{weightkappa})].

For the prediction of halo/galaxy clustering, 
we further need the projected number density of halo, $\bar{n}_{\rm h}$, and its redshift distribution per unit area, $n_{\rm h}(z)$  We estimate these quantities with halo mass-function $n(M,z)$ [see Eq.~(\ref{MF}) and (\ref{st})] through:
\begin{align}
\bar{n}_{\rm h}=\int_0^\infty dz~n_{\rm h}(z)=\int_0^\infty dz\,\frac{\chi^2(z)}{H(z)}\int_{M_{\rm min}}^\infty dM~n(M,z),
\label{nlnl}
\end{align}
where the quantities $\chi$ and $M_{\rm min}$ are respectively the comoving radial distance and the minimum mass of observed halos. In this paper, we set $M_{\rm min}$ to $10^{13}~h^{-1}M_{\odot}$.

\subsection{Angular power spectra and angular bispectra\label{sec:angpabis}}

In imaging surveys, statistical quantities relevant for cosmology are angular-power spectra and bispectra of halo/galaxy clustering and weak-gravitational lensing. In the flat-sky limit, these statistical quantities are defined as
\begin{align}
&\left\langle \Delta_{\rm a}(\bm{\ell}_1)\Delta_{\rm b}(\bm{\ell}_2)\right\rangle\equiv
(2\pi)^2\delta_{\rm D}^{(2)}(\bm{\ell}_1+\bm{\ell}_2)\,\,C_{\rm ab}(\ell_1),
\label{power2}\\
&\frac{1}{3}\Bigl[\left\langle \Delta_{\rm a}(\bm{\ell}_1)\Delta_{\rm b}(\bm{\ell}_2)\Delta_{\rm c}(\bm{\ell}_3)\right\rangle+\mbox{2 perms}\,\,(\bm{\ell}_1\leftrightarrow \bm{\ell}_2\leftrightarrow\bm{\ell}_3)\,\Bigr]
\nonumber\\
&\qquad\qquad\quad\equiv (2\pi)^2\delta_{\rm D}^{(2)}(\bm{\ell}_1+\bm{\ell}_2+\bm{\ell}_3)B_{\rm abc}(\ell_1,\ell_2,\ell_3),
\label{bis2}
\end{align}
with $\delta_{\rm D}^{(2)}$ being two-dimensional Dirac's delta function. 
The quantity $\Delta_{\rm a}$ 
is the projected field, and the subscripts $a$, $b$, $c$ imply either a halo/galaxy number-density fluctuation $\Delta_{\rm h}$ or weak-lensing convergence $\kappa$. These are related to the three-dimensional fluctuations through: 
\begin{align}
\Delta^{(2)}_{\rm h}(\bm{\theta})&=\int_0^\infty dz~ W_{\rm h}(z)\delta^{(3)}_{\rm h}(\chi(z)\bm{\theta},z),\label{delh}\\
\kappa(\bm{\theta})&=\int_0^\infty dz~ W_\kappa(z)\delta^{(3)}_{\rm m}(\chi(z)\bm{\theta},z),\label{delk}
\end{align}
with the weight functions $W_{\rm a}$ given by
\begin{align}
W_{\rm h}(z)&=\frac{n_{\rm h}(z)}{\bar{n}_{\rm h}}  ,
\\
W_\kappa\left(z\right)&=\frac{4\pi G\rho_{\rm m}\left(z\right)a^2\left(z\right)}{H\left(z\right)\bar{n}_{\rm s}}\int_z^\infty dz'~n_{\rm s}(z')\frac{\left(\chi\left(z'\right)-\chi\left(z\right)\right)\chi\left(z\right)}{\chi\left(z'\right)} .\label{weightkappa}
\end{align}
Here, $\rho_{\rm m}$ is the mean-mass density.

Substituting 
the Fourier-transform of the two-dimensional density field, $\delta^{(2)}_{\rm a}\left(\bm{\ell}\right)=\int d\bm{\theta} \delta^{(2)}_{\rm a}\left(\bm{\theta}\right)e^{-i\bm{\ell}\cdot\bm{\theta}}$, into 
Eq.(\ref{power2}) and (\ref{bis2}), the expressions for \angpab~are obtained. 
Employing the Limber approximation \cite{1954ApJ...119..655L} valid in the flat-sky limit, we have
\begin{align}
&C_{\rm ab}(\ell)=\int dz\frac{H^2(z)}{\chi^2\left(z\right)} W_{\rm a}\left(z\right)W_{\rm b}\left(z\right) \mathcal{P}_{\rm XY}\left(\frac{\ell}{\chi\left(z\right)};\,z\right),\label{plimber2}\\
&B_{\rm abc}(\bm{\ell}_1,\bm{\ell}_2,\bm{\ell}_3)=\int dz\frac{H^2(z)}{\chi^4\left(z\right)} W_{\rm a}\left(z\right)W_{\rm b}\left(z\right)W_{\rm c}\left(z\right) 
\nonumber\\
&\qquad\qquad\qquad\qquad\times
\mathcal{B}_{\rm XYZ}\left(\frac{\bm{\ell}_1}{\chi\left(z\right)},\frac{\bm{\ell}_2}{\chi\left(z\right)},\frac{\bm{\ell}_3}{\chi\left(z\right)};\,z\right),\label{blimber}
\end{align}
The right-hand side of these equations represents the projection of the three-dimensional power spectra and bispectra of the halo/galaxy clustering and matter distribution, whose expressions are given in Sec.\ref{sec:ipt}. Note that the Limber approximation becomes invalid at the large-angular scales or lower multipoles, and our treatment with flat-sky limit may not be adequate. However, if observed redshift range is wide enough, the Limber approximation is still valid even at low multipoles \cite{2008PhRvD..78l3506L}. Since we mainly examine the cases with single-redshift bin without tomography, our treatment would be still appropriate. The impact of tomographic technique will be discussed in Sec.\ref{sec:discuss}.

\subsection{Error covariance \label{sec:covmat}}

The covariances of \angpab~describe the statistical uncertainty of their 
measurements for a given survey, and their magnitude is closely related to the statistical significance of the detection and/or constraints on the 
primordial non-Gaussianity.  
The two major contributions to the covariances are the cosmic variance arising from a finite number of modes determined by the survey area, and the shot noise due to a discreteness of galaxy samples. In what follows, assuming the Gaussian covariances, we estimate the signal-to-noise ratio for \angpab, and derive the expected constraints on the primordial non-Gaussian parameters. 
The assumption for Gaussianity for the covariance matrices would be validated at $\ell\lesssim 200$ \cite{2013MNRAS.429..344K}, where the non-linear evolution of the matter density field is rather mild, and thus the gravity-induced non-Gaussian contribution is suppressed. Further, a large primordial non-Gaussianity is not allowed by the current observations. Then, the covariance matrices of \angpab~ for a given set of multipole bins $(\ell_i,\,\ell_j,\,\cdots)$ 
are expressed as \cite{2013MNRAS.429..344K}
\begin{align}
&{\rm Cov}[C_{\rm ab}(\ell_i),C_{\rm a'b'}(\ell_j)]
=\frac{\delta^{\rm K}_{\bm{\ell}_i+\bm{\ell}_j}}{N_{\rm pairs}(\ell_i)}
\nonumber\\
&\times\Bigl[
\left\{C_{\rm aa'}(\ell_i)+N_{\rm aa'}\right\}\,\left\{C_{\rm bb'}(\ell_i)+N_{\rm bb'}\right\}+\mbox{perms}\,({\rm a'\leftrightarrow b'})\,\Bigr],
\label{covpower}
\\
&{\rm Cov}[B_{\rm abc}(\ell_i,\ell_j,\ell_k),B_{\rm a'b'c'}(\ell_l,\ell_m,\ell_n)]
=\frac{1}{9}\frac{\Omega_{\rm s}}{N_{\rm trip}(\ell_i,\ell_j,\ell_k)}
\nonumber\\
&\times\Biggl[\Bigl\{\left(C_{\rm aa'}(\ell_i)+N_{\rm aa'}\right)\left(C_{\rm bb'}(\ell_j)+N_{\rm bb'}\right)\left(C_{\rm cc'}(\ell_k)+N_{\rm cc'}\right)
\nonumber\\
&\quad\times\left(\delta^{\rm K}_{\bm{\ell}_i+\bm{\ell}_l}\delta^{\rm K}_{\bm{\ell}_j+\bm{\ell}_m}\delta^{\rm K}_{\bm{\ell}_k+\bm{\ell}_n}+\mbox{5 perms}\,\,(\ell_l\leftrightarrow \ell_m\leftrightarrow \ell_n)\right)
\nonumber\\
&\quad+\mbox{2 perms}\,\,({\rm a'\leftrightarrow b'\leftrightarrow c'})\Bigr\}
+\mbox{2 perms}\,\,({\rm a\leftrightarrow b\leftrightarrow c})\Biggr],
\label{covB}
\end{align}
with the quantity $N_{\rm ab}$ being the shot-noise contribution given by
\begin{align}
&N_{\rm ab}=\left\{
\begin{array}{ll}
{\displaystyle \frac{1}{\bar{n}_{\rm h}}}& (\mbox{ab}=\mbox{hh}),
\\
\\
{\displaystyle \frac{\sigma_\gamma}{\bar{n}_{\rm s}}} & (\mbox{ab}=\kappa\kappa),
\\
\\
{\displaystyle 0} & (\mbox{otherwise}).
\end{array}
\right.
\label{covbis}
\end{align}
Here, $\delta^{\rm K}_{\bm{\ell}_i+\bm{\ell}_j}$ is the Kronecker delta, and     
the quantity $\sigma_\gamma$ represents the dispersion of the intrinsic shape noise to which we set $\sigma_\gamma=0.3$ \cite{2013PhR...530...87W}. 
The  $N_{\rm pair}(\ell_i)$ is the number of independent pairs for two vectors $\bm{\ell}$ and $-\bm{\ell}$ within the $i$-th multipole bin, i.e.,  
$\ell_i-\Delta \ell_i/2\leq|\bm{\ell}|\leq\ell_i+\Delta \ell_i/2$. 
The $N_{\rm trip}(\ell_i,\ell_j,\ell_k)$ is the number of independent triplets for three vectors forming the triangular configuration 
respectively within the $i$-, $j$-, and $k$-th bins. 
In the limit $\ell_i\gg\ell_{\rm f}$, where $\ell_{\rm f}\equiv2\pi/\Omega_{\rm s}$ is the minimum multipoles determined by the fundamental Fourier mode with $\Omega_{\rm s}$ being the survey area in units of steradian, we obtain the analytical expressions for $N_{\rm pair}$ and $N_{\rm trip}$ \cite{2013MNRAS.429..344K}: 
\begin{align}
&N_{\rm pair}(\ell_i)\simeq\frac{2\pi \ell_i\Delta \ell_i}{\ell_{\rm f}^2},\label{npair}\\
&N_{\rm trip}(\ell_i,\ell_j,\ell_k)\simeq 2\frac{(2\pi \ell_i\Delta \ell_i)(\ell_j\Delta\varphi_{12}\Delta \ell_j)}{\ell_{\rm f}^4}
\end{align}
with the angle $\Delta\varphi_{12}$ given by
\begin{align}
\Delta\varphi_{12}(\ell_i,\ell_j,\ell_k)&\simeq (\sin{\varphi_{12}})^{-1}\frac{\ell_k\Delta \ell_k}{\ell_i\ell_j}
\nonumber
\\
&=\frac{2\ell_k\Delta \ell_k}{\sqrt{2\ell_i^2\ell_j^2+2\ell_i^2\ell_k^2+2\ell_j^2\ell_k^2-\ell_i^4-\ell_j^4-\ell_k^4}}. 
\label{ntrip}
\end{align}
Note that the width of the $i$-th multipole bin, $\Delta \ell_i$, should be larger than the minimum multipole, i.e., $\Delta \ell_i> \ell_{\rm f}$. 

\section{Signal-to-noise ratio\label{sec:SN}}

In this section, based on the formulas presented in Sec.~\ref{sec:powerandbi} and \ref{sec:obs}, we first compute the angular-power spectra and bispectra 
in Sec.~\ref{angpabpho}, specifically focusing on HSC survey. 
The feasibility to break parameter degeneracy between higher-order non-Gaussian parameters is also discussed. Then, in Sec.~\ref{SNangpab}, 
the signal-to-noise ratio of \angpab~is estimated for three representative surveys. 

\subsection{Angular-power spectra and bispectra from imaging surveys\label{angpabpho}}

For illustrative purpose, let us focus on the HSC survey, and 
present the expected results for the measurements of both power spectrum and bispectrum. To elucidate how the primordial non-Gaussianity changes the shape and amplitude of power spectrum and bispectrum, it is convenient to decompose the contributions in \angpab~into the gravity-induced part and non-Gaussian part:
\begin{align}
&C_{\rm ab}(\ell)=C_{\rm grav}(\ell)+f_{\rm NL}\,C_{f_{\rm NL}}(\ell)+g_{\rm NL}\,C_{g_{\rm NL}}(\ell)+\tau_{\rm NL}\,C_{\tau_{\rm NL}}(\ell),
\label{eq:power3D_2}\\
&B_{\rm abc}\left(\ell_1,\ell_2,\ell_3\right)=B_{\rm grav}\left(\ell_1,\ell_2,\ell_3\right)+f_{\rm NL}\,B_{f_{\rm NL}}\left(\ell_1,\ell_2,\ell_3\right)
\nonumber\\
&\qquad\quad\quad+g_{\rm NL}\,B_{g_{\rm NL}}\left(\ell_1,\ell_2,\ell_3\right)+\tau_{\rm NL}\,B_{\tau_{\rm NL}}\left(\ell_1,\ell_2,\ell_3\right).
\label{bisbis3D}
\end{align}
Here, just for brevity, the subscripts $ab$ are omitted at the right-hand-side.  Each term is evaluated by substituting the three-dimensional power spectra and bispectra [Eqs.~(\ref{Ptreee})-(\ref{2-loopbi})] into Eqs.~(\ref{plimber2}) and (\ref{blimber}).

\begin{figure*}[htbp] 
\vspace{-3.8cm}
\hspace*{-0.5cm}
  \includegraphics[width=190mm]{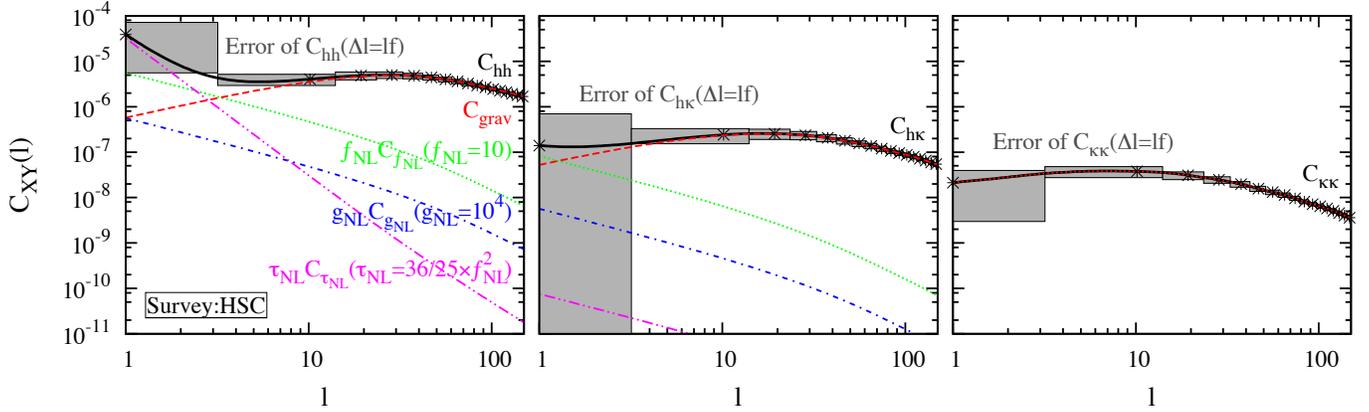}
\vspace{-5.7cm}
\caption{Angular auto- and cross-power spectra of halo clustering and 
weak lensing as function of multipoles $\ell$ in the presence of primordial non-Gaussianity; $C_{\rm hh}$ (left), $C_{\rm h\kappa}$ (middle) and $C_{\kappa\kappa}$ (right). Assuming the survey parameters of HSC, and the non-Gaussian parameters of $f_{\rm NL}=10$, $g_{\rm NL}=10^4$, and $\tau_{\rm NL}=(36/25) f_{\rm NL}^2$, the expected signals for power spectra are estimated based on iPT. In each panel, 
black solid lines represent the total amplitude of the power spectrum,  
while the lines with different colors indicate the partial contributions to the power spectra defined in Eq.~(\ref{eq:power3D_2});  
$C_{\rm grav}$ (red), $f_{\rm NL}C_{f_{\rm NL}}$ (green), $g_{\rm NL}C_{g_{\rm NL}}$ (blue), and $\tau_{\rm NL}C_{\tau_{\rm NL}}$ (magenta). 
The grey boxes represent the expected errors on the total amplitude of power spectra, averaged over each bin. Here, the width of the bin is set to $\Delta \ell = \ell_{\rm f}$. \label{sigerrofpower}}
\end{figure*}

Fig.~\ref{sigerrofpower} shows 
the auto-power spectra of halos ({\it left}) and 
lensing convergence ({\it right}) and their cross spectrum ({\it middle}),  
assuming the non-Gaussian parameters of $f_{\rm NL}=10$, $g_{\rm NL}=10^4$, and $\tau_{\rm NL}=(36/25) f_{\rm NL}^2$. 
The statistical errors expected from the HSC survey are 
depicted as shaded boxes. Note that in computing the power spectra, we use the 
expressions in the large-scale limit presented in Appendix~\ref{sec:largelim}. 
Fig.~\ref{sigerrofpower} shows that with the currently allowed values of the non-Gaussianity, the contributions of the primordial non-Gaussianity can dominate the auto-power spectrum of halo/galaxy clustering at $\ell\lesssim 5$, where the cosmic variance is the dominant source for statistical error. 
Thus, the surveys with wide survey area are preferable and HSC survey has a potential to  tightly constrain the primordial non-Gaussianity. However, 
the large-scale limit of angular power spectra asymptotically scales as
\begin{align}
&C_{\rm hh}(\ell):~ C_{\rm grav}\propto \ell,~~ C_{f_{\rm NL}}\propto \ell^{-1},~~ C_{g_{\rm NL}}\propto \ell^{-1},~~ C_{\tau_{\rm NL}}\propto \ell^{-3},\label{ldepphh}\\
&C_{\rm h\kappa}(\ell):~ C_{\rm grav}\propto \ell,~~ C_{f_{\rm NL}}\propto \ell^{-1},~~ C_{g_{\rm NL}}\propto \ell^{-1},~~ C_{\tau_{\rm NL}}\propto \ell^{-1},
\label{ldepphk}\\
&C_{\rm \kappa\kappa}(\ell):~ C_{\rm grav}\propto \ell,
\label{ldeppkk}
\end{align}
which are derived based on the formulas in Sec.\ref{sec:powerandbi}, assuming the scale-invariant primordial spectrum, i.e., $P_{\Phi}(k)\propto k^{-3}$. 
These asymptotic behaviors are at least valid  at $\ell\gtrsim 30$. This implies that even when the primordial non-Gaussian contribution becomes prominent, one cannot separately detect and/or constrain both $f_{\rm NL}$ and $g_{\rm NL}$ from the power spectra. As we will see in next section, due to this parameter degeneracy, the marginalized constraints on $f_{\rm NL}$ and $g_{\rm NL}$ from the Fisher matrix analysis are substantially reduced (see Table\ref{HSCfis}).

\begin{figure*}[htbp]
\hspace*{-0.5cm}
\includegraphics[width=165mm]{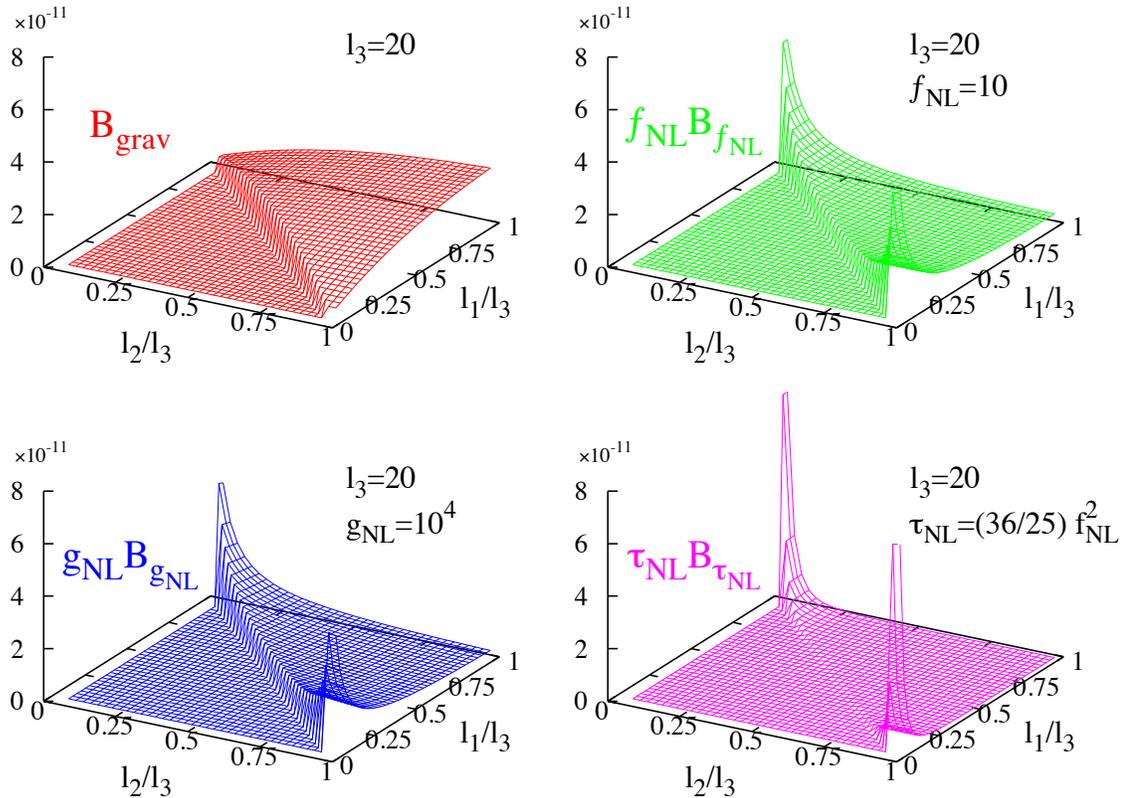}
\vspace{-1.cm}
 \caption{Shape of the angular bispectrum of halos, $B_{\rm hhh}$, in the presence of primordial non-Gaussianity. Fixing the third multipole $\ell_3$ to $20$, and setting the non-Gaussian parameters to $f_{\rm NL}=10$, $g_{\rm NL}=10^4$, and $\tau_{\rm NL}=(36/25) f_{\rm NL}^2$, the resultant amplitude of bispectrum are plotted as function of $\ell_2/\ell_3$ and $\ell_1/\ell_3$.  
The four different panels separately show each contribution to the halo bispectrum, defined in Eq.~(\ref{bisbis3D}); $B_{\rm grav}$ (top left), $f_{\rm NL}B_{f_{\rm NL}}$ (top right), $g_{\rm NL}B_{g_{\rm NL}}$ (bottom left), and $\tau_{\rm NL}B_{\tau_{\rm NL}}$ (top right). 
  \label{3DBhhh}}
\end{figure*}
\begin{figure*}[htbp]
\hspace*{1cm}
  \includegraphics[width=180mm]{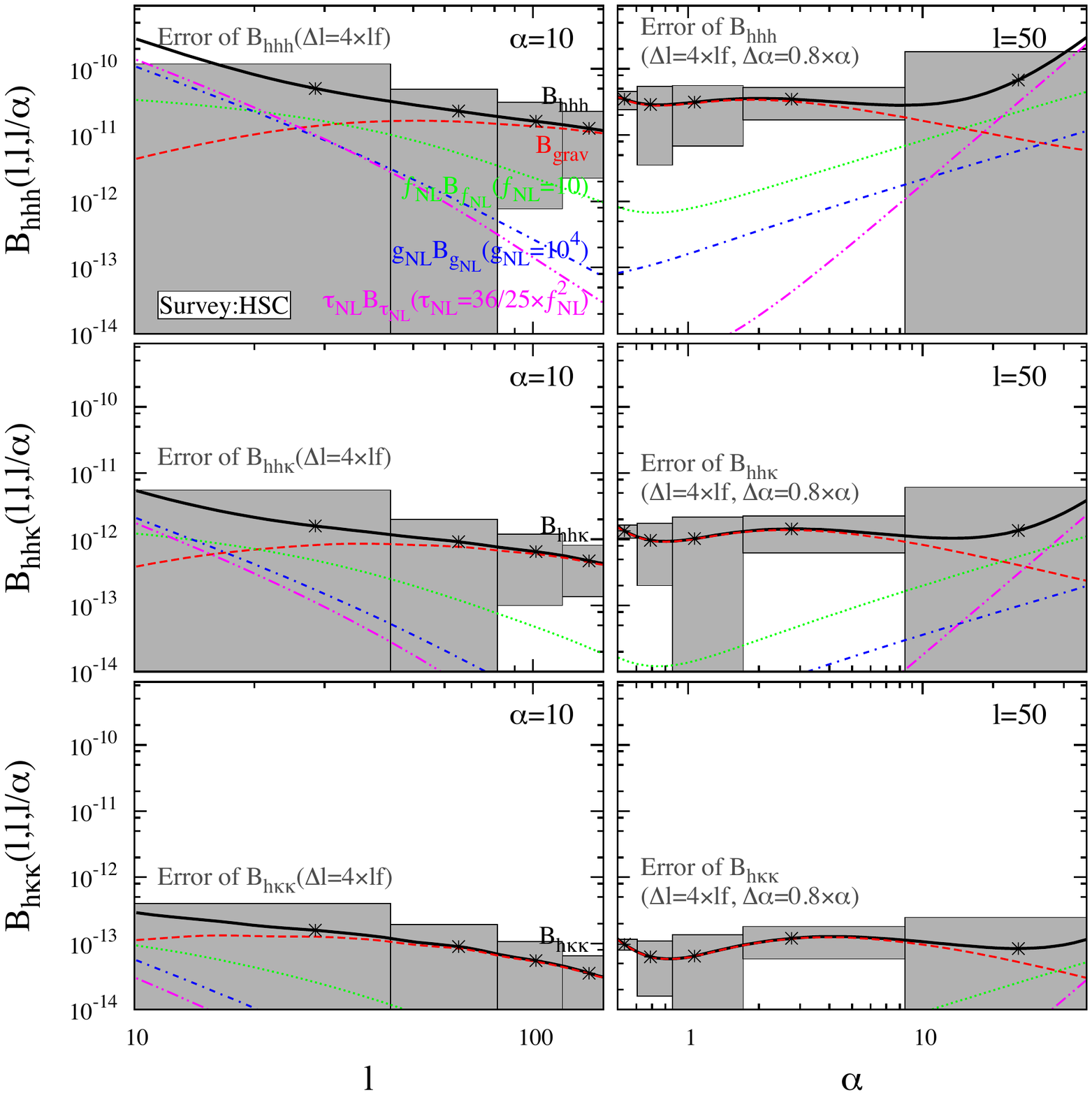}
\vspace{-0.5cm}
\caption{Scale dependence of the angular auto-/cross-bispectra of halo density and weak-lensing fields in the squeezed configuration, $B(\ell,\ell,\ell/\alpha)$. The auto-bispectrum $B_{\rm hhh}$ (top), cross-bispectra $B_{\rm hh\kappa}$ (middle) and $B_{\rm h\kappa\kappa}$ (bottom) are computed with iPT, and the results are plotted assuming the non-Gaussian parameters of $f_{\rm NL}=10$, $g_{\rm NL}=10^4$, and $\tau_{\rm NL}=(36/25) f_{\rm NL}^2$. 
Left panels show the bispectrum amplitude fixing the parameter $\alpha$ to $10$, and plot the results as function of multipole $\ell$. On the other hand, right panels show the bispectrum fixing the multipole 
$\ell$ to $50$, and plot the results as function of $\alpha$.  
In each panel, 
black solid lines represent the total amplitude of the bispectrum,  
while the lines with different colors indicate the partial contributions to the bispectrum defined in Eq.~(\ref{bisbis3D}). 
$B_{\rm grav}$ (red), $f_{\rm NL}B_{f_{\rm NL}}$ (green), $g_{\rm NL}B_{g_{\rm NL}}$ (blue), and $\tau_{\rm NL}B_{\tau_{\rm NL}}$ (magenta). 
The grey boxes represent the expected errors on the total amplitude of power spectra, averaged over each bin. Here, the width of the bin is set to $\Delta \ell = 4\ell_{\rm f}$ and $\delta\alpha=0.8\alpha$. 
\label{BzuB}}
\end{figure*}

On the other hand, the primordial non-Gaussian contributions 
to the angular bispectra have different asymptotic behaviors. 
Fig.~\ref{3DBhhh} shows the shape dependence of the halo bispectra $B_{\rm hhh}$,  plotted against $\ell_1/\ell_3$ and $\ell_2/\ell_3$ with fixed $\ell_3=20$.  
The dominant contributions from primordial non-Gaussianity
appear in the squeezed limit,  i.e., $\ell_3=\ell_2\gg \ell_1~{\rm or}~\ell_3=\ell_1\gg \ell_2$, and the primordial non-Gaussian terms have distinct shape and amplitude dependence from the gravity-induced non-Gaussian term, $B_{\rm grav}$. 
Although $B_{f_{\rm NL}}$ and $B_{g_{\rm NL}}$ apparently look very similar, 
their contributions possess different scale-dependent behaviors. 
To clarify this, we consider the isosceles configuration given by $\ell\equiv\ell_1=\ell_2=\ell_3/\alpha$. A large $\alpha$ implies the squeezed shape. 
Then, for small $\ell$ and large $\alpha$, the asymptotic behaviors of each contribution become\footnote{In deriving Eq.~(\ref{ldepb1}), we assume that the dominant contributions to the terms $B_{\rm grav},~ B_{f_{\rm NL}},~ B_{g_{\rm NL}}$ and $B_{\tau_{\rm NL}}$ respectively come from $\mathcal{B}^{\rm tree}_{\rm grav}$ [Eq.~(\ref{treegrav})], $\mathcal{B}^{\rm tree}_{\rm bis}$ [Eq.~(\ref{Btreebis-L})], $\mathcal{B}^{\rm 1-loop}_{g_{\rm NL}}$ [Eq.~(\ref{1-loopgnlB})] and $\mathcal{B}^{\rm 2-loop}_{\tau_{\rm NL}}$ [Eq.~(\ref{2-looptaunlB})].}
\begin{align}
B_{\rm hhh}(\ell,\ell,\ell/\alpha)&:~ B_{\rm grav}\propto \ell^2\alpha^0,~~ B_{f_{\rm NL}}\propto \ell^0\alpha^1,~~ B_{g_{\rm NL}}\propto \ell^{-2}\alpha^1,
\nonumber\\
&~~~ B_{\tau_{\rm NL}}\propto \ell^{-2}\alpha^3.\label{ldepb1}
\end{align}
Note that we assume $P_{\Phi}(k)\propto k^{-3}$. The shape dependence of each contribution is manifestly different between each other. The cross bispectra, $B_{\rm hh\kappa}(\ell,\ell,\ell/\alpha)$ and $B_{\rm h\kappa\kappa}(\ell,\ell,\ell/\alpha)$, are also shown to have similar behaviors (see Fig.~\ref{BzuB}).

Fig.~\ref{BzuB} plots the scale-dependence of the auto- and cross-bispectra of the isosceles configuration of $(\ell, \ell,\ell/\alpha)$. Left and right panels respectively show the bispectra as function of multipoles and $\alpha$, fixing the arguments to $\alpha=10$ and $\ell=50$. Statistical errors are also depicted as gray boxes, assuming the HSC survey. Fig.~\ref{BzuB} shows that the auto bispectra $B_{\rm hhh}$ is most sensitive to 
the primordial non-Gaussian contributions, which become dominant at low multipoles and large $\alpha$. In particular, 
with the non-Gaussian parameters of $f_{\rm NL}=10$, $g_{\rm NL}=10^4$, and $\tau_{\rm NL}=(36/25) f_{\rm NL}^2$, the term $B_{\tau_{\rm NL}}$ is found to exceed other non-Gaussian contributions. Although the statistical error of the squeezed configuration ($\alpha\gg1$) at lower multipoles looks still large, we will show in next subsection that summing up various configurations lead to a sufficiently high signal-to-noise ratio, with which the signature of primordial non-Gaussianity becomes measurable.

\subsection{Signal-to-noise ratio\label{SNangpab}}

To quantify the detectability of the signal,  
we here define the signal-to-noise ratio (SNR) for power spectra and bispectra. 
Combining all auto- and cross-power spectra, the SNR for power spectra is :  
\begin{align}
&\left(\frac{S}{N}\right)^2_{{\rm all}~ P}\equiv \sum_{\ell_{\rm min}\le \ell_i,\ell_j\le \ell_{\rm max}}\bm{C}(\ell_i)\left[{\rm Cov}^{P}\right]^{-1}_{ij}\bm{C}(\ell_j),\label{SNpower}
\end{align}
with the quantities $\bm{C}$  and ${\rm Cov}^{P}$ respectively given by
\begin{align}
&\bm{C}(\ell)=\begin{pmatrix}
C_{\rm hh}(\ell)  \\
C_{\rm h\kappa}(\ell)  \end{pmatrix},
\\
&{\rm Cov}_{ij}^P=
\begin{pmatrix}
{\rm Cov}[C_{\rm hh}(\ell_i),C_{\rm hh}(\ell_j)]& {\rm Cov}[C_{\rm hh}(\ell_i),C_{\rm h\kappa}(\ell_j)]\\
{\rm Cov}[C_{\rm h\kappa}(\ell_i),C_{\rm hh}(\ell_j)]& {\rm Cov}[C_{\rm hh}(\ell_i),C_{\rm \kappa\kappa}(\ell_j)]\end{pmatrix}.
\label{ppmatrix}
\end{align}
The subscripts $i$ and $j$ run over the multipole bins within the range $[\ell_{\rm min},\,\ell_{\rm max}]$. The quantity $\left[{\rm Cov}^{P}\right]^{-1}_{ij}$ is the inverse of the power spectra covariance matrix defined in Eq.~(\ref{ppmatrix}). Similarly, the SNR for bispectra is defined as follows: 
\begin{align}
&\left(\frac{S}{N}\right)^2_{{\rm all}~ B}\equiv \sum_{\ell_{\rm min}\le\{\ell_i\},\{\ell_j\}\le \ell_{\rm max}}\bm{B}_i\left[{\rm Cov}^{B}\right]^{-1}_{ij}\bm{B}_j,\label{SNbis}
\end{align}
with 
\begin{widetext}
\begin{align}
&\bm{B}_i=\begin{pmatrix}
(B_{\rm hhh})_i \\
(B_{\rm hh\kappa})_i  \\
(B_{\rm h\kappa\kappa})_i  \end{pmatrix},
\qquad
\bm{{\rm Cov}}_{ij}^B=\begin{pmatrix}
{\rm Cov}[(B_{\rm hhh})_i,(B_{\rm hhh})_j]& {\rm Cov}[(B_{\rm hhh})_i,(B_{\rm hh\kappa})_j]& {\rm Cov}[(B_{\rm hhh})_i,(B_{\rm h\kappa\kappa})_j]\\
{\rm Cov}[(B_{\rm hh\kappa})_i,(B_{\rm hhh})_j]& {\rm Cov}[(B_{\rm hh\kappa})_i,(B_{\rm hh\kappa})_j]& {\rm Cov}[(B_{\rm hh\kappa})_i,(B_{\rm h\kappa\kappa})_j]\\
{\rm Cov}[(B_{\rm h\kappa\kappa})_i,(B_{\rm hhh})_j]& {\rm Cov}[(B_{\rm h\kappa\kappa})_i,(B_{\rm hh\kappa})_j]& {\rm Cov}[(B_{\rm h\kappa\kappa})_i,(B_{\rm h\kappa\kappa})_j] \end{pmatrix}\label{covB2}.
\end{align}
\end{widetext}
Again, subscripts $i$ and $j$ run over all possible triangle configurations, and we include all the triangle configurations whose side length are within the range $[\ell_{\rm min},\,\ell_{\rm max}]$. We set the minimum multipole to $\ell_{\rm min}=\ell_{\rm f}=2\pi/\sqrt{\Omega_{\rm s}}$.  

\begin{figure*}[htbp]
\vspace*{-5.cm}
  \includegraphics[width=180mm]{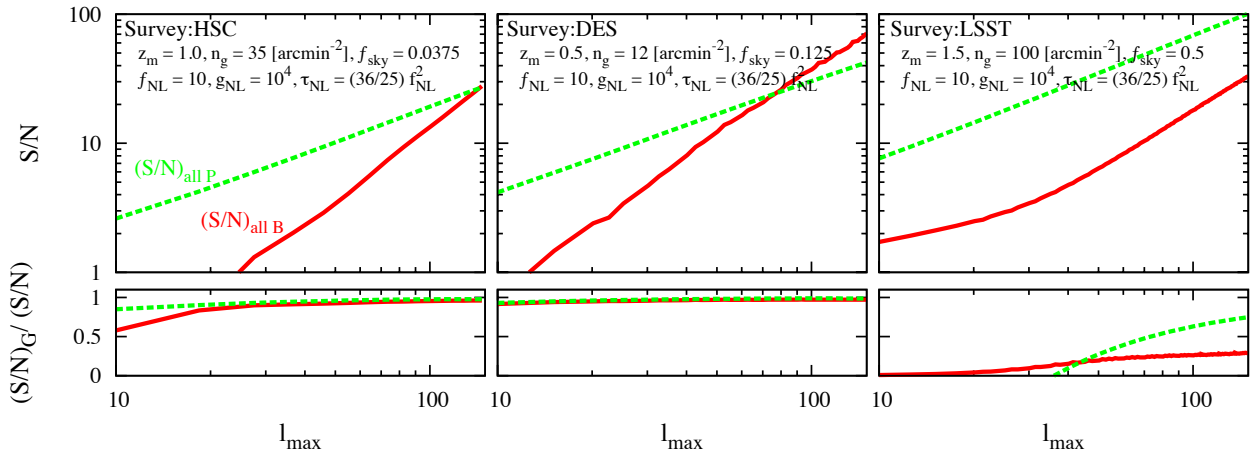}
\vspace*{-2.7cm}
  \caption{Signal-to-noise ratio for angular power spectra and bispectra combining both the halo clustering and weak lensing measurements. Based on the parameters specified in Table \ref{ngsmfsky} and assuming the non-Gaussian parameters of $f_{\rm NL}=10$, $g_{\rm NL}=10^4$, and $\tau_{\rm NL}=(36/25) f_{\rm NL}^2$, the resultant signal-to-noise ratio for 
three representative surveys of HSC (left), DES (center) and LSST (right) 
are shown as function of the maximum multipole $\ell_{\rm max}$. The upper panels are the estimated results of $(S/N)_{{\rm all}~P}$ (dashed) and $(S/N)_{{\rm all}~B}$ (solid) [see Eqs.~(\ref{SNpower}) and (\ref{SNbis}) for definitions]. The bottom panels show the ratio of signal-to-ratio with and without primordial non-Gaussianity, defined by $(S/N)_{\rm G}/(S/N)$. 
  \label{SN_HSC}}
\end{figure*}

In top panels of Fig.~\ref{SN_HSC}, the resultant values of SNR, 
$(S/N)_{{\rm all}~ P}$ (green dashed) and $(S/N)_{{\rm all}~ B}$ (red solid), are plotted as function of maximum multipoles for representative three surveys; HSC (left), DES (middle) and LSST (right). In all surveys, both of 
$(S/N)_{{\rm all}~ P}$ and $(S/N)_{{\rm all}~ B}$ exceed 
$10$ at $\ell_{\rm max}\gtrsim50$, and  
the feasibility to measure power spectra and bispectra is 
ensured at high statistical significance. A notable point is that 
the SNR of bispectra rapidly increases with maximum multipoles, and 
for HSC and DES, it eventually exceeds that of power spectra 
at $\ell\gtrsim100$. This is because the number of possible configurations of bispectra, given by $\sum_{\ell_{\rm min}\le\{\ell_i\},\{\ell_j\}\le \ell_{\rm max}}N_{\rm trip}(\ell_i,\ell_j,\ell_{\rm max})$, grows faster than that of power spectra $N_{\rm pair}(\ell_{\rm max})$.

On the other hand, bottom panels of Fig.~\ref{SN_HSC} plot the ratio of SNR with and without primordial non-Gaussianity, defined by $(S/N)_{\rm G}/(S/N)$. Here, we assume $f_{\rm NL}=10$, $g_{\rm NL}=10^4$, and $\tau_{\rm NL}=(36/25)f^2_{\rm NL}$ in evaluating the denominator of this ratio. The results 
that the ratio is close to unity for HSC (left) and DES (middle) 
indicate that 
the signals of both power spectra and bispectra are mostly dominated by the gravity-induced terms, and the primordial non-Gaussian contributions only amounts few percents even at $\ell_{\rm max}=150$. By contrast, for LSST, the estimated values of the ratio $(S/N)_{\rm G}/(S/N)$ become $0.7$ and $0.3$ for power spectra and bispectra, respectively, and the primordial non-Gaussian contributions amount $\sim 30\%$ and $\sim 70\%$ for the total SNR. 
This is because LSST will cover the high-redshift range where nonlinearity of the gravity is weak. Further, sky coverage of LSST is large enough and the effect of \png~can be enhanced. Although the resultant SNRs of bispectra are almost the same for three surveys, LSST is potentially much more powerful to detect primordial non-Gaussianity among others.

\section{Forecast constraints on primordial non-Gaussianity\label{sec:fifi}}

\begin{figure*}[tb]
\hspace*{-1.2cm}
   \includegraphics[width=7.6cm]{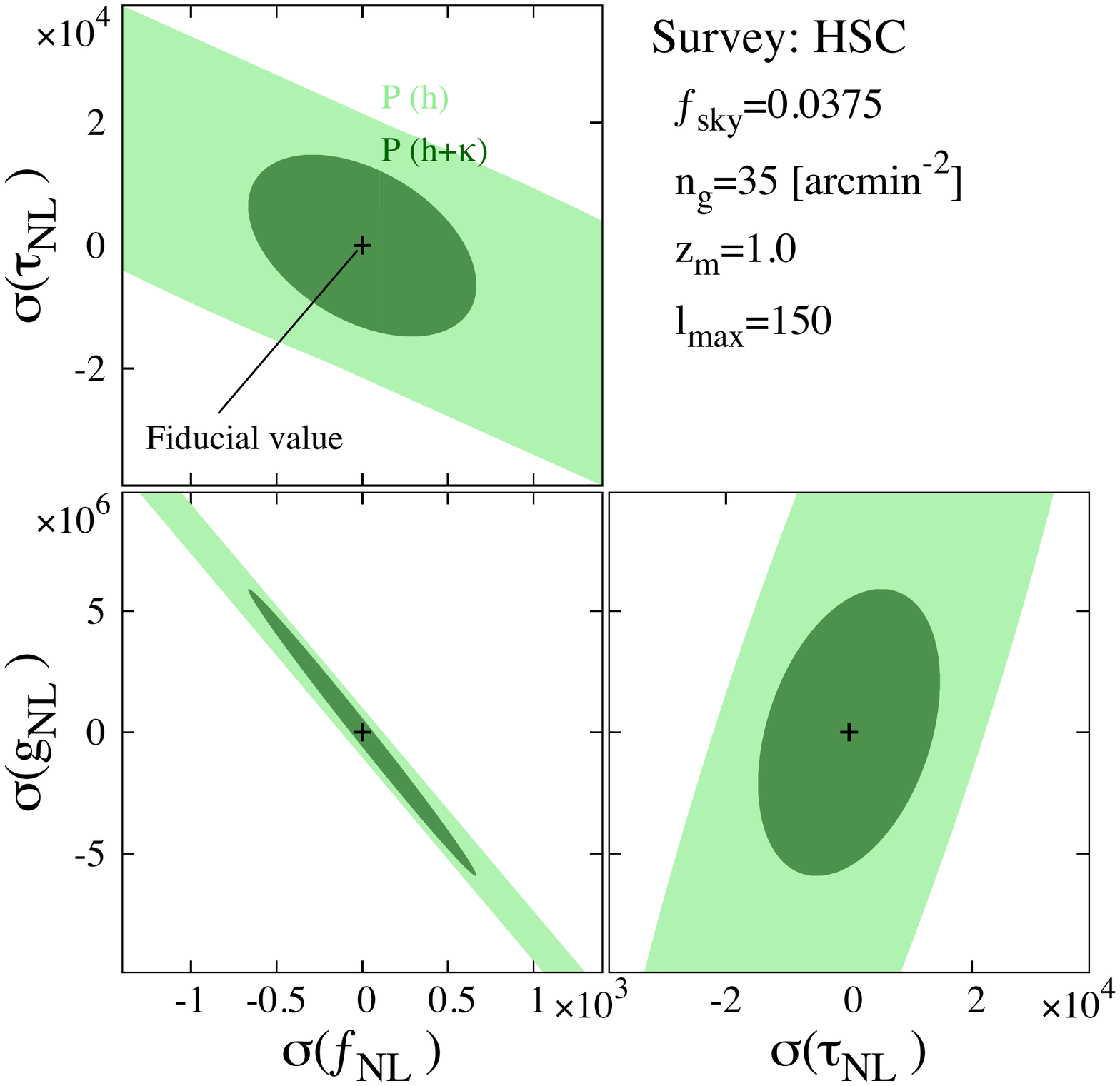}
\hspace*{-2.2cm}
  \includegraphics[width=7.6cm]{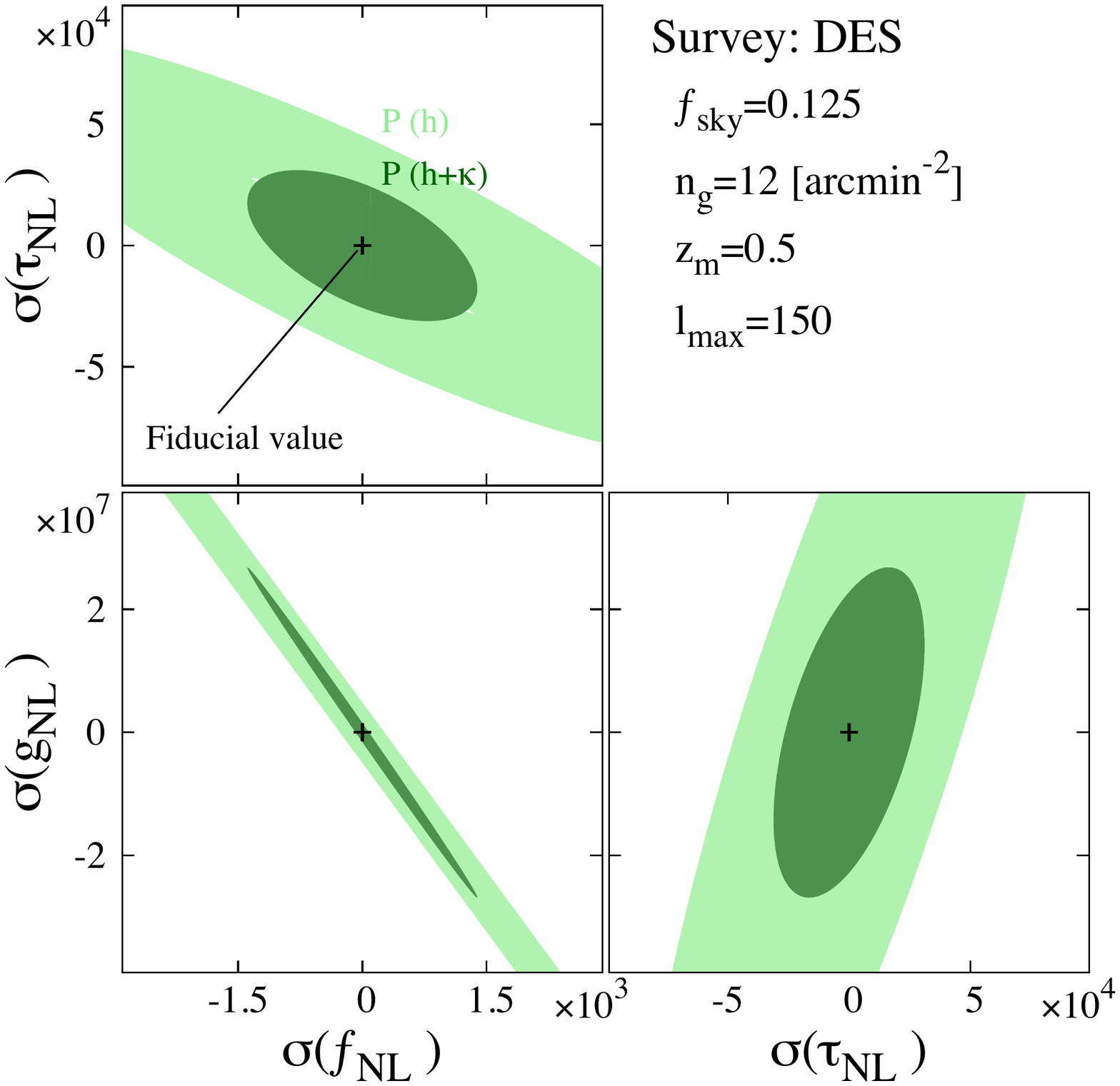}
\hspace*{-2.2cm}
  \includegraphics[width=7.6cm]{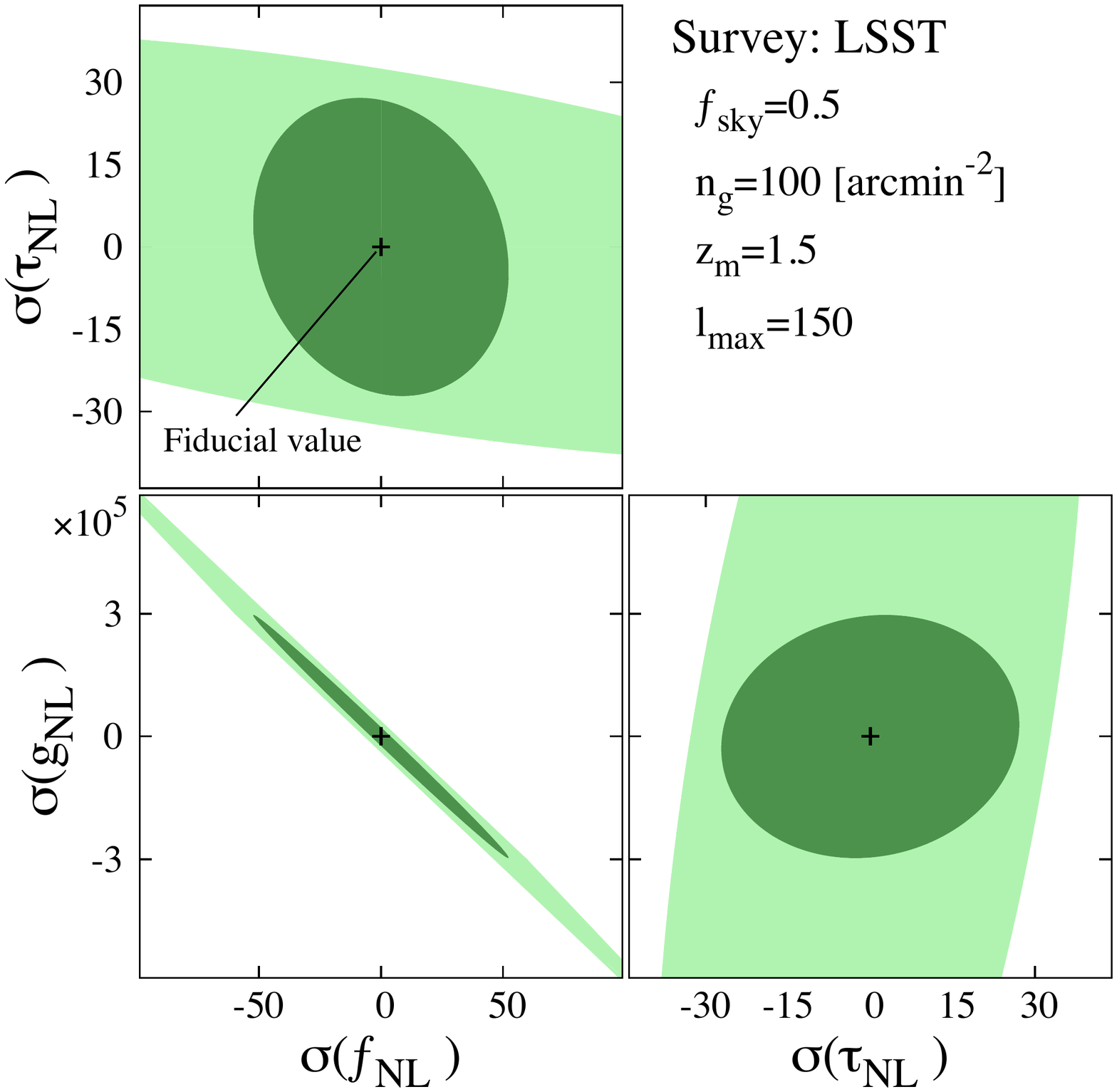}
  \caption{Forecast results of primordial non-Gaussian parameter for HSC (left), DES (middle) and LSST (right). In each panel, the marginalized $1\sigma$ error contours on $\tau_{\rm NL}-f_{\rm NL}$ (top left), $g_{\rm NL}-f_{\rm NL}$ (bottom left) and $g_{\rm NL}-\tau_{\rm NL}$ (bottom right) planes are plotted. Faint green contours indicate the expected constraints derived from the angular power spectrum of halo clustering, while the dark green contours represent the results when we add the cross correlation between halo clustering and weak lensing. For clarity, the plotted range of the error contours is changed in each panel. }
  \label{PPhh_HSC_DES_LSST}
  \vspace*{0.5cm}
\hspace*{-1.2cm}
   \includegraphics[width=7.6cm]{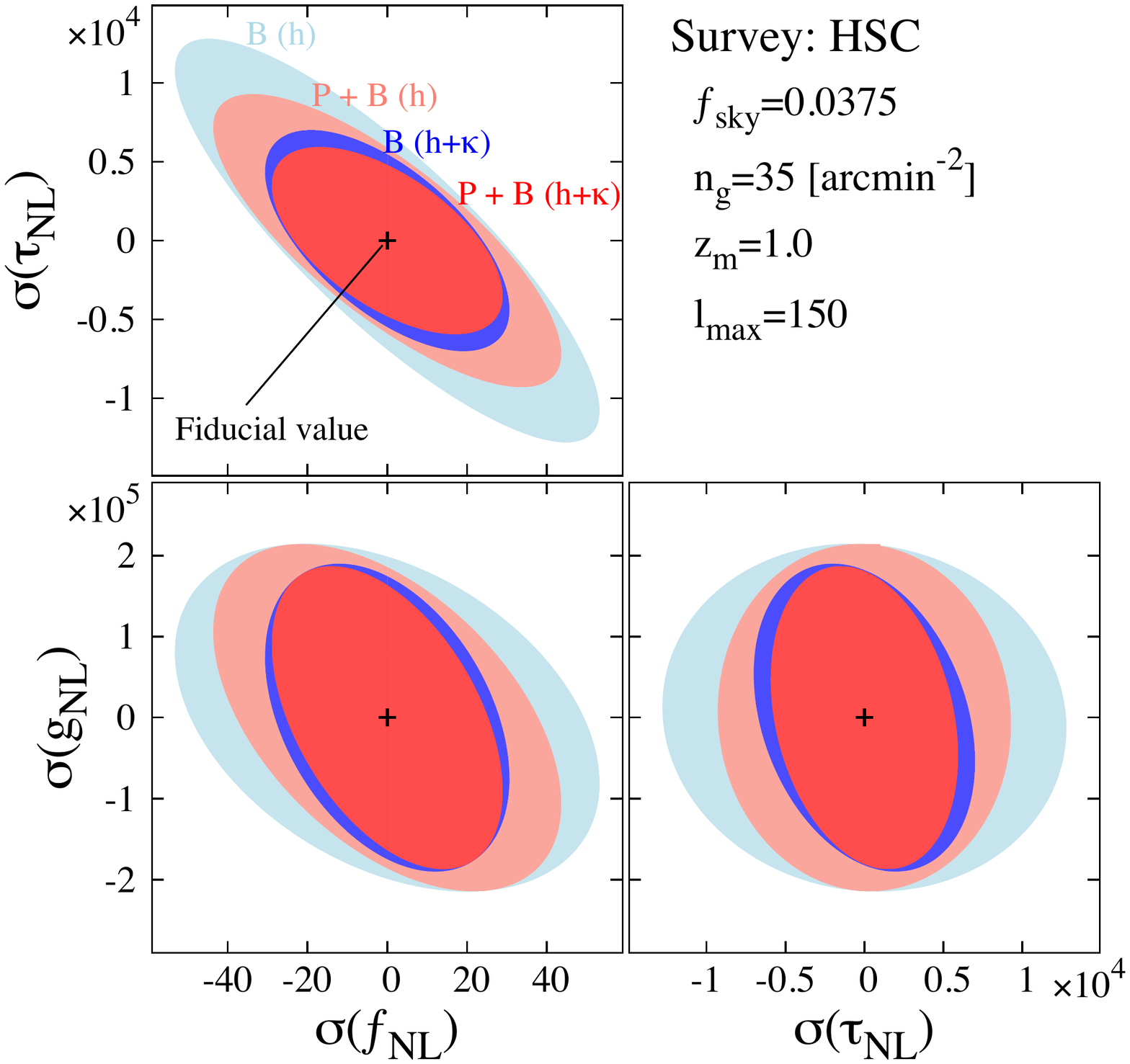}
\hspace*{-2.2cm}
   \includegraphics[width=7.6cm]{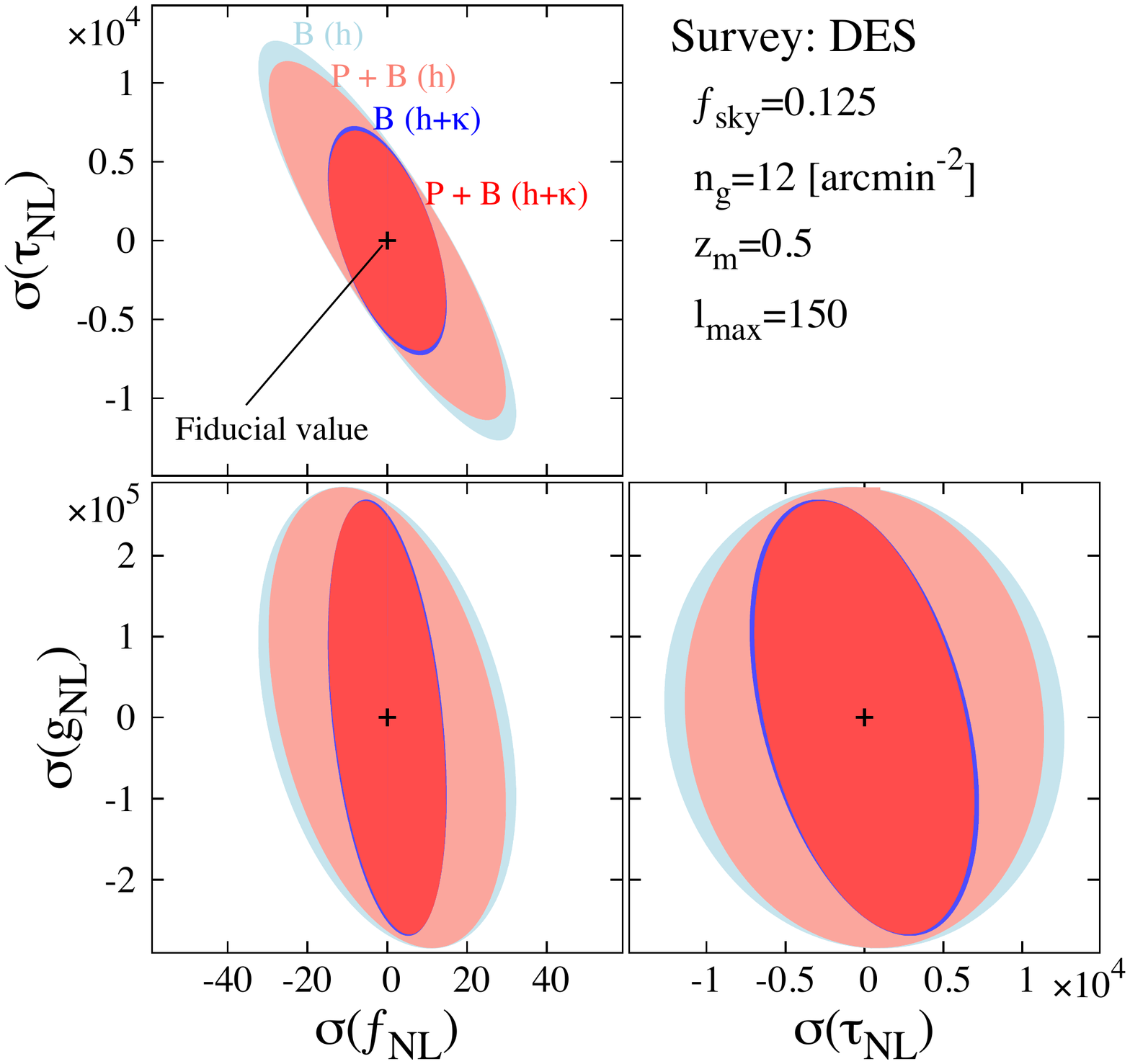}
\hspace*{-2.2cm}
   \includegraphics[width=7.6cm]{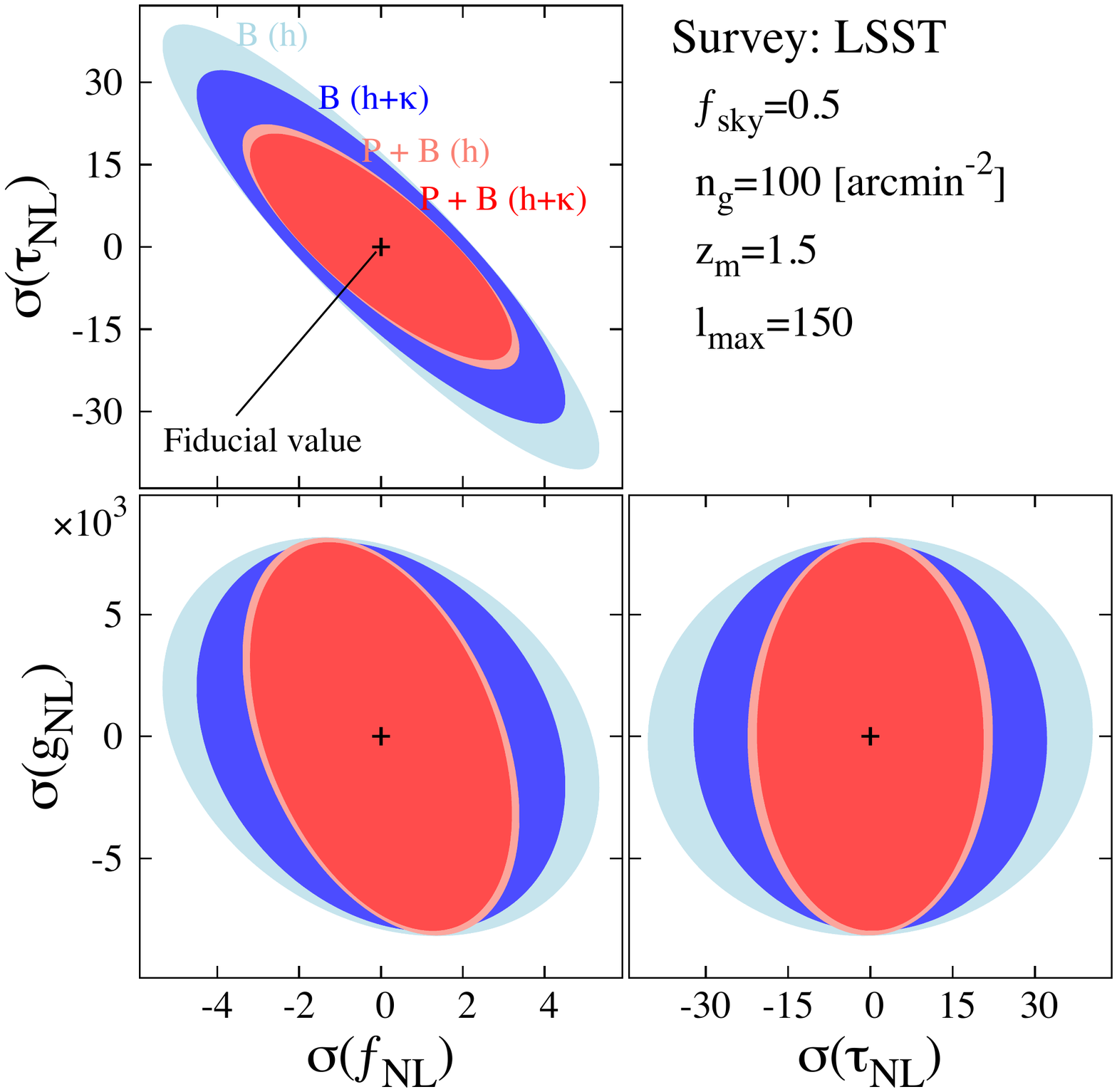}
  \caption{Forecast results of primordial non-Gaussian parameters by HSC (left), DES (middle), and LSST (right). In each panel, marginalized $1\sigma$ error contours on $\tau_{\rm NL}-f_{\rm NL}$ (top left), $g_{\rm NL}-f_{\rm NL}$ (bottom left) and $g_{\rm NL}-\tau_{\rm NL}$ (bottom right) planes are shown. Light blue contours represent the constraints from the auto-angular bispectrum of halo/galaxy clustering, while the blue contours indicate the constraints when we add cross-angular bispectrum between halos and  weak lensing. On the other hand, the light red contours show the results combining both the auto-power spectrum and bispectrum of halos. The red contours indicates the case combining all the auto- and cross-power spectra and bispectra of halos and weak lensing. For clarity, the plotted range of the error contours is changed in each panel. }
  \label{2dfisherfgt_HSC_DES_LSST}
\end{figure*}

  \begin{table*}[tb]
\begin{ruledtabular}
  \begin{tabular}{c|lll|l}
     &HSC&DES&LSST& ~current CMB~~~\\ \hline
    $\sigma (\fnl)$&$19\ (9.2)$&$9.8\ (5.2)$&$2.1\ (0.89)$&$(5.1)$\\ 
    $\sigma (\gnl)$&$1.2\times 10^5\ (7.1\times 10^4)$&$1.8\times 10^5\ (1.0\times 10^5)$&$5.3\times 10^3\ (3.8\times 10^3)$&$(1.4\times 10^5)$\\ 
    $\sigma (\tnl)$&$3.9\times 10^3\  (2.1\times 10^3)$&$4.6\times 10^3\ (2.5\times 10^3)$&$14\ (6.2)$&$(1.4\times 10^3)$\\
 \end{tabular}
  \caption{Forecast results of marginalized (un-marginalized) $1\sigma$ errors on primordial non-Gaussian parameters for HSC, DES, and LSST. The results are compared with with the single-parameter constraint derived from CMB measurement by Planck \cite{2013arXiv1303.5084P}.}
  \label{hhhdddlll}
  \end{ruledtabular}
  \end{table*}

On the basis of Fisher-matrix formalism, we are now in position to derive the expected constraints on the higher-order non-Gaussian parameters for three representative surveys. 

Provided the theoretical template of angular-power spectra and bispectra parametrized by a set of free parameters $\bm{p}$, the Fisher matrix for the parameters $\bm{p}$ are defined by 
\begin{align}
F_{\alpha\beta}&=F_{\alpha\beta}^P+F_{\alpha\beta}^B;
\label{eq:Fisher_total}\\
F_{\alpha\beta}^P&=\left.\sum^{\ell_{\rm max}}_{\ell_i=\ell_{\rm min}}\frac{\partial\bm{C}(\ell_i,\bm{p})}{\partial p_\alpha}(\bm{{\rm Cov}}^P)_{ij}^{-1}\frac{\partial\bm{C}(\ell_j,\bm{p})}{\partial p_\beta}\right|_{\bm{p}=\bm{p}_0},
\label{eq:Fisher_P}\\
F_{\alpha\beta}^B&=\left.\sum^{\ell_{\rm max}}_{\ell_i=\ell_{\rm min}}\frac{\partial\bm{B}_{i}(\bm{p})}{\partial p_\alpha}(\bm{{\rm Cov}}^B)_{ij}^{-1}\frac{\partial\bm{B}_{j}(\bm{p})}{\partial p_\beta}\right|_{\bm{p}=\bm{p}_0},
\label{eq:Fisher_B}
\end{align}
where $\bm{p}_0$ is a set of fiducial cosmological parameters. 
Here, the quantities $F_{\alpha\beta}^P$ and $F_{\alpha\beta}^B$ 
are the Fisher matrices for power spectra and bispectra, respectively, 
which are assumed to be statistically independent. 
Fisher matrix $F_{\alpha\beta}$ quantifies the statistical uncertainty for the parameters $\bm{p}$ that are determined by observations, and the $1\sigma$ (68\%\,C.L.) statistical error on the single parameter $p_\alpha$ is given by $(1/F_{\alpha\alpha})^{1/2}$. Also, the $1\sigma$ statistical error marginalized over other parameters is expressed as $([F]^{-1}_{\alpha\alpha})^{1/2}$ with $[F]^{-1}$ being the inverse of Fisher matrix.

\begin{table*}[tb]
\begin{ruledtabular}
  \begin{tabular}{|c|c|c|c|c|c|}
     HSC&$C_{\rm hh}$&$C_{\rm hh}+C_{\rm \kappa h}$&$B_{\rm hhh}$&$B_{\rm hhh}+B_{\rm hh\kappa}+B_{\rm h\kappa\kappa}$&$C_{\rm hh}+C_{\rm \kappa h}+$\\ 
      survey&&&&&$B_{\rm hhh}+B_{\rm hh\kappa}+B_{\rm h\kappa\kappa}$\\ \hline
    $\sigma (\fnl)$&$2.3\times 10^3\ (30)$&$4.2\times 10^2\ (21)$&$32\ (11)$&$20\ (10)$&$19\ (9.2)$\\ 
    $\sigma (\gnl)$&$2.0\times 10^7\ (2.8\times 10^5)$&$3.4\times 10^6\ (1.7\times 10^5)$&$1.3\times 10^5\ (7.9\times 10^4)$&$1.3\times 10^5\ (7.8\times 10^4)$&$1.2\times 10^5\ (7.1\times 10^4)$\\ 
    $\sigma (\tnl)$&$2.9\times 10^4\ (7.5\times 10^3)$&$8.3\times10^3\ (4.0\times 10^3)$&$7.1\times 10^3\ (2.5\times 10^3)$&$4.6\times 10^3\ (2.4\times 10^3)$&$3.9\times 10^3\  (2.1\times 10^3)$\\ 
    \end{tabular}
  \caption{Forecast results of marginalized (un-marginalized) $1\sigma$ errors on primordial non-Gaussian parameters for HSC.
  \label{HSCfis}}
  \end{ruledtabular}

\vspace*{0.5cm}

\begin{ruledtabular}
  \begin{tabular}{|c|c|c|c|c|c|}
     DES &$C_{\rm hh}$&$C_{\rm hh}+C_{\rm \kappa h}$&$B_{\rm hhh}$&$B_{\rm hhh}+B_{\rm hh\kappa}+B_{\rm h\kappa\kappa}$&$C_{\rm hh}+C_{\rm \kappa h}+$\\ 
     \quad &&&&&$B_{\rm hhh}+B_{\rm hh\kappa}+B_{\rm h\kappa\kappa}$\\ \hline
    $\sigma (\fnl)$&$3.1\times 10^3\ (45)$&$9.2\times 10^2\ (30)$&$21\ (6.3)$&$9.9\ (5.3)$&$9.8\ (5.2)$\\ 
    $\sigma (\gnl)$&$6.2\times 10^7\ (9.1\times 10^5)$&$1.8\times 10^7\ (6.0\times 10^5)$&$1.9\times 10^5\ (1.1\times 10^5)$&$1.8\times 10^5\ (1.1\times 10^5)$&$1.8\times 10^5\ (1.0\times 10^5)$\\ 
    $\sigma (\tnl)$&$5.7\times 10^4\ (1.8\times 10^4)$&$2.0\times10^4\ (1.0\times 10^4)$&$8.4\times 10^3\ (2.6\times 10^3)$&$4.8\times 10^3\ (2.5\times 10^3)$&$4.6\times 10^3\ (2.5\times 10^3)$\\ 
    \end{tabular}
  \caption{Forecast results of marginalized (un-marginalized) $1\sigma$ errors on primordial non-Gaussian parameters for DES.
  \label{DESfis}}
  \end{ruledtabular}

\vspace*{0.5cm}
\begin{ruledtabular}
  \begin{tabular}{|c|c|c|c|c|c|}
     LSST&$C_{\rm hh}$&$C_{\rm hh}+C_{\rm \kappa h}$&$B_{\rm hhh}$&$B_{\rm hhh}+B_{\rm hh\kappa}+B_{\rm h\kappa\kappa}$&$C_{\rm hh}+C_{\rm \kappa h}+$\\ 
     \quad&&&&&$B_{\rm hhh}+B_{\rm hh\kappa}+B_{\rm h\kappa\kappa}$\\ \hline
    $\sigma (\fnl)$&$2.1\times 10^2\ (1.4)$&$35\ (1.3)$&$3.5\ (1.3)$&$3.0\ (1.3)$&$2.1\ (0.89)$\\ 
    $\sigma (\gnl)$&$1.2\times 10^6\ (8.1\times 10^3)$&$2.0\times 10^5\ (7.2\times 10^3)$&$5.4\times 10^3\ (4.5\times 10^3)$&$5.3\times 10^3\ (4.4\times 10^3)$&$5.3\times 10^3\ (3.8\times 10^3)$\\ 
    $\sigma (\tnl)$&$26\ (9.6)$&$18\ (8.4)$&$27\ (10)$&$21\ (9.3)$&$14\ (6.2)$
\\ 
    \end{tabular}
  \caption{Forecast result of marginalized (un-marginalized) $1\sigma$ errors on primordial non-Gaussian parameters for LSST.
  \label{LSSTfis}}
  \end{ruledtabular}
  \end{table*} 
 
Here, for free parameters $\bm{p}$, we consider the three non-Gaussian parameters, i.e., $\bm{p}=(f_{\rm NL}, \,g_{\rm NL}, \,\tau_{\rm NL})$, with the fiducial values of $\bm{p}_0=(0,0,0)$. We do not marginalize the uncertainty in the halo bias properties, since our PT treatment with iPT completely specifies the halo clustering for a given minimum halo mass, $M_{\rm min}$, which we set $10^{13}\,h^{-1}$\,M$_\odot$. This may be a rather optimistic assumption, however, 
our primary purpose is to explore the feasibility to test single-sourced consistency relation by simultaneously constraining multiple non-Gaussian parameters. 
Since the properties of the clustering bias is expected to be observationally determined at the relatively small scales, where no notable effect of the primordial non-Gaussianity appears, there would be no serious parameter degeneracy with non-Gaussian parameters. As we will see below, the choice of our minimum halo mass does not significantly affect the expected constraints on non-Gaussianity. A forecast study in more practical situation will be considered elsewhere. 

In evaluating the Fisher matrix, 
we set the minimum multipole to $\ell_{\rm min}=\ell_{\rm f}=2\pi/\sqrt{\Omega_{\rm s}}$, while the maximum multipole $\ell_{\rm max}$ is set to $150$, adopting the Gaussian covariances in 
Eqs.~(\ref{covpower}) and (\ref{covB}). Note that increasing $\ell_{\rm max}$ may give a tighter constraint on the non-Gaussian parameters, as 
naively inferred from Fig.~\ref{SN_HSC}.  
However, a big impact on the statistical analysis may come from the gravity-induced non-Gaussian contributions to the error covariances, for which we do not consider. In this sense, our results presented below may be regarded as a conservative estimate, and a possibility to further improve the constraints needs to be investigated.

The results of Fisher matrix analysis are summarized in 
Figs.~\ref{PPhh_HSC_DES_LSST} and \ref{2dfisherfgt_HSC_DES_LSST},  
and Tables \ref{HSCfis}-\ref{LSSTfis} for 
each survey. In Table \ref{hhhdddlll}, the forecast results from three surveys are compared to the current constraints from CMB.

Let us first look at the forecast results from the power spectra.  
Figs.~\ref{PPhh_HSC_DES_LSST} show the two-dimensional error contours on $(f_{\rm NL},g_{\rm NL})$ (bottom left), $(f_{\rm NL}, \tau_{\rm NL})$ and $(\tau_{\rm NL},g_{\rm NL})$ from the power spectrum data. As we expect from the asymptotic scale-dependence in Eq.~(\ref{ldepphk}), the forecast constraints from halo clustering data alone exhibit a large degeneracy between each parameter (see also Ref.~\cite{Yokoyama:2011qr}), and the marginalized constraints are substantially degraded, compared to the single-parameter cases (see second column in Tables~\ref{HSCfis}-\ref{LSSTfis}). Adding the weak lensing power spectra partly breaks the parameter degeneracy, and constraints becomes tightened, especially for the parameter $\tau_{\rm NL}$\footnote{The expected errors on $f_{\rm NL}$ for HSC are somewhat degraded compared to the previous study by Refs.~\cite{2011PhRvD..83l3514N,2012PhRvD..85d3518T}. These differences mainly come from the choice of minimum multipole, $\ell_{\rm min}$. While the previous study adopted $\ell_{\rm min}=2$, we conservatively set it to the fundamental mode, i.e.,  $\ell_{\rm min}=\ell_{\rm f}$, which roughly corresponds to $9$ for HSC. }. The constraint on $f_{\rm NL}$ and $g_{\rm NL}$ is also improved, however, degeneracy still remains between these two parameters, and the resultant values of the marginalized error are still larger than those of the un-marginalized constraint by more than one order of magnitude. This is clearly seen in the third columns of Tables~\ref{HSCfis}-\ref{LSSTfis}.

\begin{figure*}[tb]
\begin{center}
\vspace*{-1.0cm}
   \includegraphics[width=160mm]{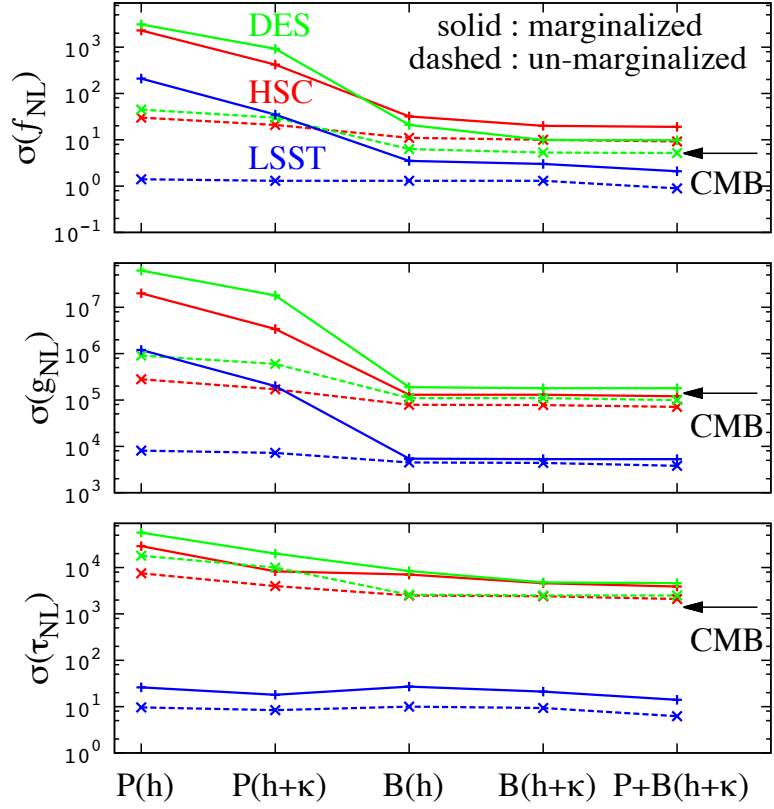}
  \end{center}
  \vspace*{-1.0cm}
  \caption{Variation of the marginalized (solid) and un-marginalized (dashed) errors on the parameters $f_{\rm NL}$ (top), $g_{\rm NL}$ (middle), and $\tau_{\rm NL}$ (bottom) with respect to what kinds of statistics are used/combined;  
the auto-power spectrum of halos $[P(h)]$, auto- and cross-power 
spectra of halos and weak lensing $[P(h+\kappa)]$, 
the auto-bispectrum of halo clustering $[B(h)]$, 
the auto- and cross-bispectra of halos and weak lensing $[B(h+\kappa)]$, and 
finally, the combination of both the power spectra and bispectra of halos and weak lensing  $[P+B(h+\kappa)]$. In each panel, 
green, red and blue lines are the results for DES, HSC, and LSST, respectively. 
For reference, horizontal arrows indicate the single-parameter constraints derived from the latest CMB measurement \cite{2013arXiv1303.5084P}. 
  \label{surv_const}}
\end{figure*}

The situation is drastically changed if we use the bispectrum data, as shown in Figs.~\ref{2dfisherfgt_HSC_DES_LSST} and \ref{surv_const}. A remarkable point is that only with the halo bispectrum, the degeneracy between three parameters are mostly broken, and this enables us to simultaneously constrain each parameter. As a result, the marginalized constraints become rather close to the un-marginalized ones (forth column in Tables~\ref{HSCfis}-\ref{LSSTfis} and Fig.~\ref{surv_const}). This is indeed expected from the asymptotic scaling relation in Eq.~(\ref{ldepb1}). Further adding the cross correlation data moderately improves the constraints, and the expected constraints on $f_{\rm NL}$ and $\tau_{\rm NL}$ are improved by a factor of $<1.5$, compared to the halo bispectrum alone. 

The results imply that a deep imaging survey is advantageous to give a tight constraint on primordial non-Gaussianity, especially for $g_{\rm NL}$ and $\tau_{\rm NL}$. As an idealistically wide and deep survey, the LSST can give tighter constraints on $f_{\rm NL}$, $g_{\rm NL}$, and $\tau_{\rm NL}$, and the expected constraints will be significantly tighter than those obtained from the CMB (see Table~\ref{hhhdddlll} and Fig.~\ref{surv_const}). Note that the constraints from the CMB summarized in 
Table~\ref{hhhdddlll} and Fig.~\ref{surv_const} are un-marginalized ones. In this respect, HSC and DES can still give a meaningful result on non-Gaussianity, and LSST would be the most sensitive non-Gaussian probe suited for testing single-sourced consistency relation. 

\begin{figure*}[tb]
\begin{center}
\hspace*{-1.2cm}
   \includegraphics[width=135mm]{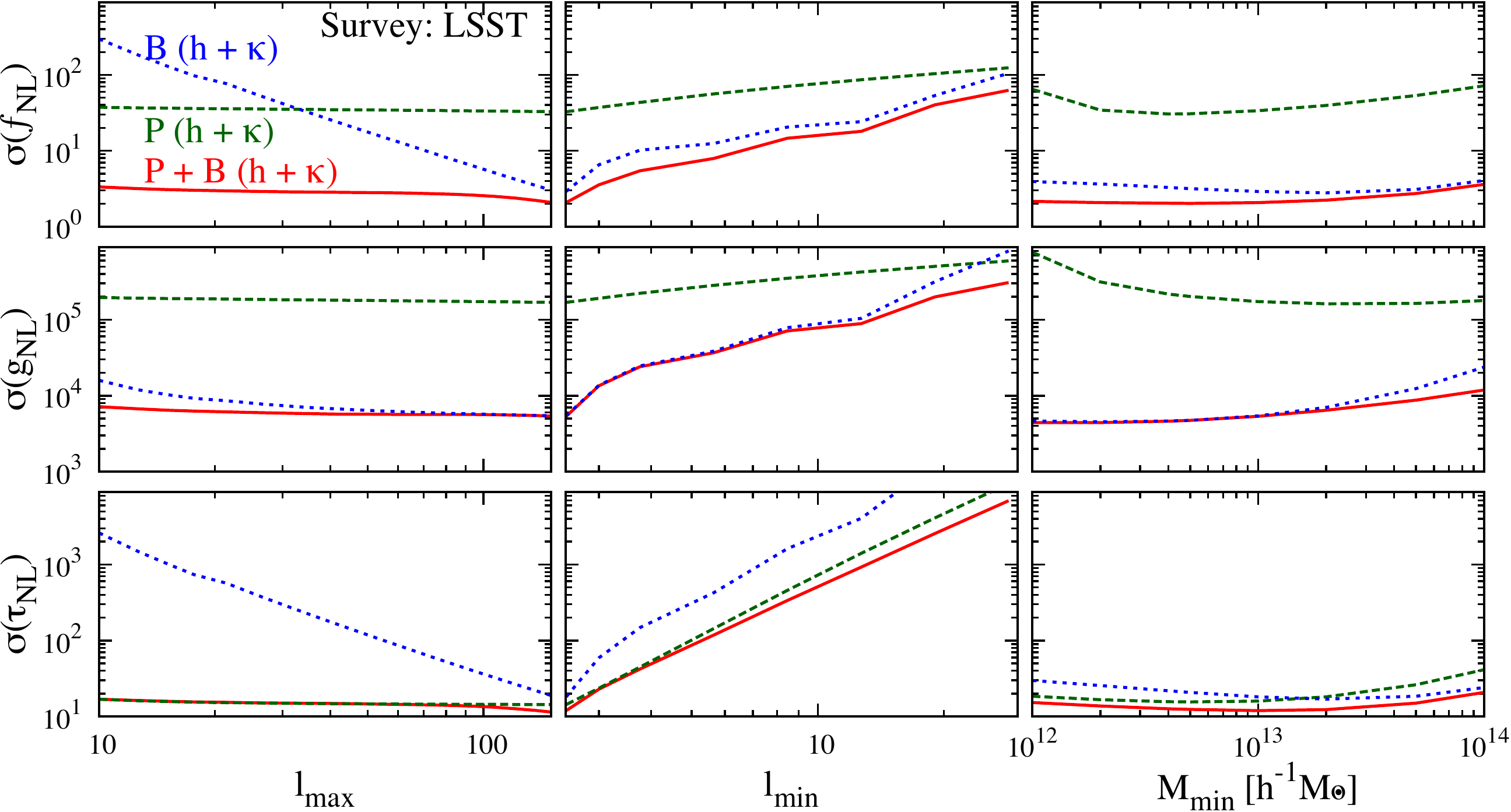}
  \end{center}
  \vspace*{-0.75cm}
  \caption{Sensitivity of the marginalized $1\sigma$ errors on $f_{\rm NL}$ (top), $g_{\rm NL}$ (center), and $\tau_{\rm NL}$ (bottom) to the parameters $\ell_{\rm max}$ (left), $\ell_{\rm min}$ (middle) and $M_{\rm min}$ (right). We here particularly consider LSST. In plotting the results as function of each parameter, the other two parameters are kept fixed at the fiducial values (i.e., $\ell_{\rm max}=150$, $\ell_{\rm min}=\ell_{\rm f}\simeq 2.5$ and  $M_{\rm min}=10^{13} \,h^{-1}$\,M$_\odot$). Red solid lines represent the constraints derived from the combinations of all auto-/cross-angular bispectra and power spectra, while the green dashed and blue dotted lines respectively indicate the results derived from the power spectra and bispectra.}
  \label{lmaxdep}
\end{figure*}

Finally, to see how the resultant constraints depend on our choice of the parameters, we focus on the LSST survey, and vary one of the minimum and maximum multipoles $(\ell_{\rm min}$ and $\ell_{\rm max})$, and the minimum halo mass $(M_{\rm min})$. The variation of the marginalized constraints on non-Gaussian parameters is summarized in Fig.~\ref{lmaxdep}.

Left panels show the dependence on $\ell_{\rm max}$. The constraints coming from the bispectra (blue dotted) continuously become improved as increasing $\ell_{\rm max}$. This is rather contrasted to the power spectrum cases in which the constraining power becomes saturated at rather low multipoles. 
The reason for the improvement basically comes from the fact that the fractional contribution of the squeezed triangles becomes dominant with $\ell_{\rm max}$. Since the local-type primordial non-Gaussianity is known to produce a large bispectrum amplitude at squeezed limit, Fig.~\ref{lmaxdep} suggests that a drastic improvement of the constraints will be obtained if we have a reliable theoretical template for bispectrum relevant to the small scales, $\ell\gtrsim 150$. On the other hand, middle panels show the dependence on $\ell_{\rm min}$. Since the present methodology makes full use of the scale-dependent properties of the halo clustering, the limiting sky coverage significantly degrades the constraining power on primordial non-Gaussianity. In particular, higher-order parameters, $g_{\rm NL}$ and $\tau_{\rm NL}$, are very sensitive to $\ell_{\rm min}$ because of the strong scale-dependence of the terms, $C_{\tau_{\rm NL}}$ and $B_{\tau_{\rm NL}}$ [see Eqs.~(\ref{ldepphh}), (\ref{ldepphk}), and (\ref{ldepb1})]. Finally, right panels plot the dependence on $M_{\rm min}$ while $\ell_{\rm min}$ and $\ell_{\rm max}$ are kept fixed. In general, a large minimum halo mass selects highly biased halos, and the resultant clustering amplitudes get increased. The number of halos is, however, decreased, and the shot-noise contribution is increased. As a result of two competing effects, the $M_{\rm min}$ dependance on the constraints becomes weak at the mass range of $10^{12}\,h^{-1}$\,M$_\odot < M_{\rm min} < 10^{14}\,h^{-1}$\,M$_\odot$.

\begin{figure*}[tb]
\hspace*{-0.5cm}
   \includegraphics[width=95mm]{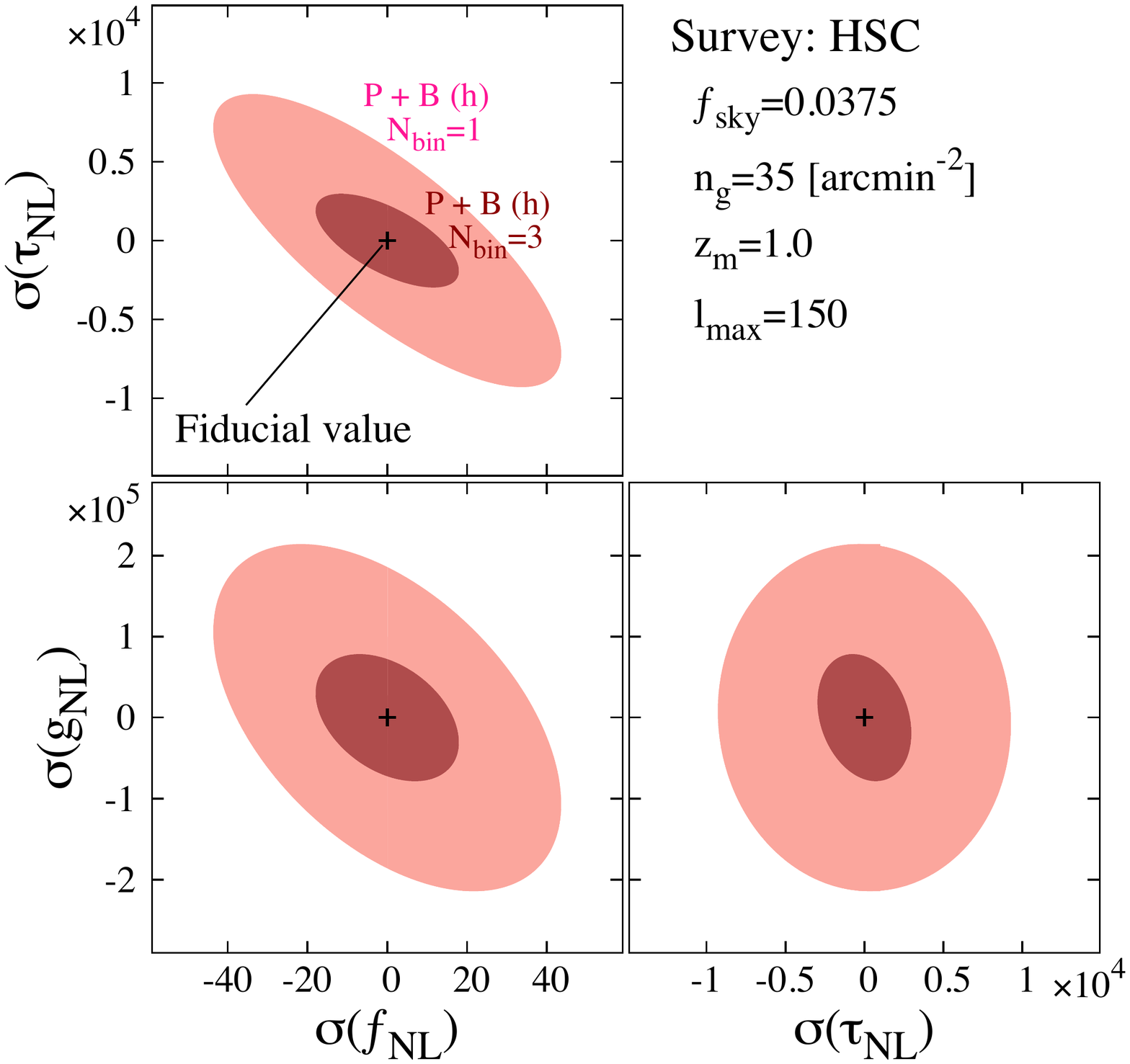}
\hspace*{-1.5cm}
   \includegraphics[width=95mm]{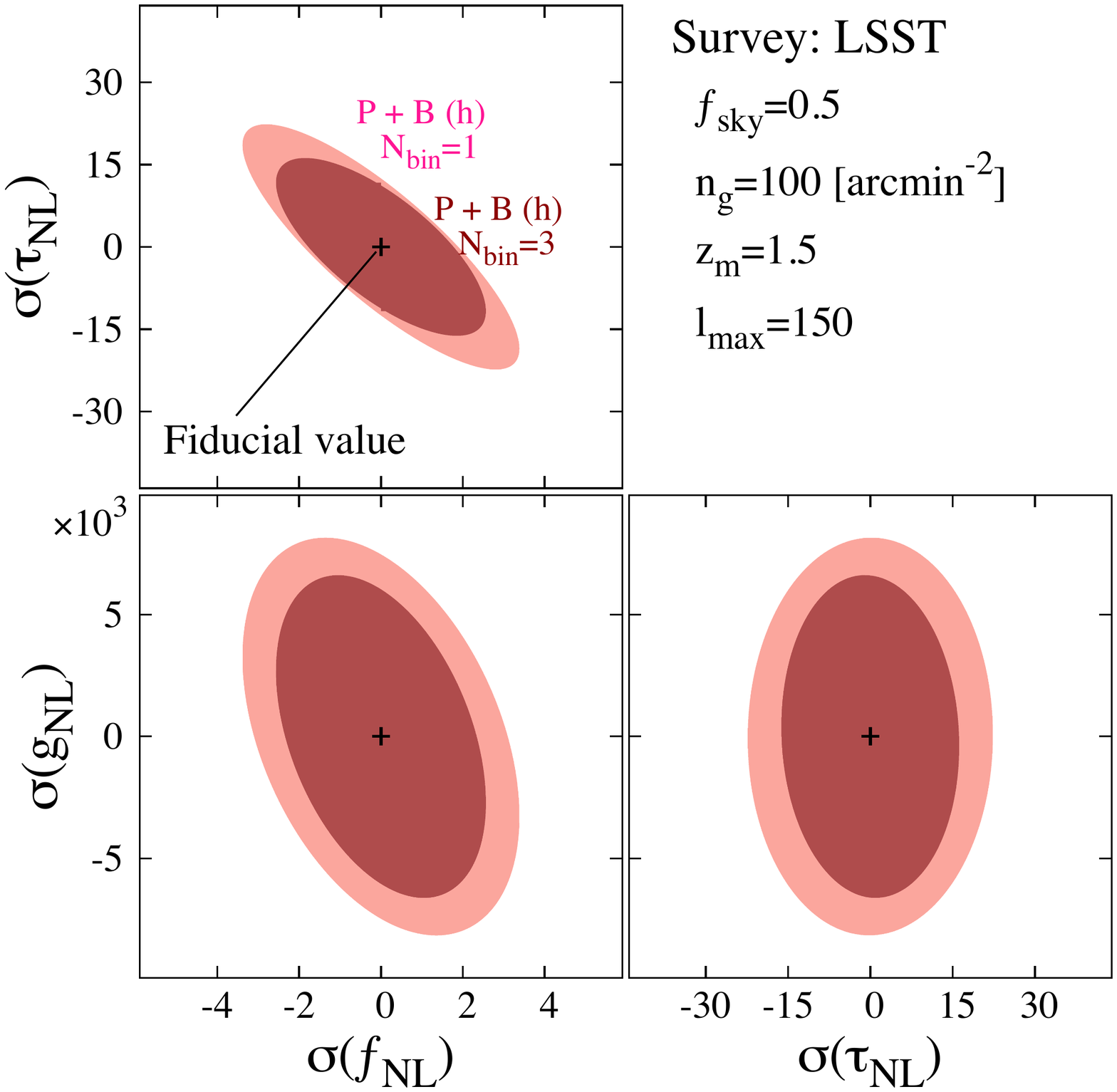}
\vspace*{-0.5cm}
  \caption{Forecast results of marginalized $1\sigma$ error contours on $\tau_{\rm NL}-f_{\rm NL}$ (top left), $g_{\rm NL}-f_{\rm NL}$(bottom left) and $g_{\rm NL}-\tau_{\rm NL}$(bottom right) planes for HSC (left) and LSST (right). In each panel, we show the constraints of the combination of the auto-angular power spectra and bispectra with $N_{\rm bin}=1$ (salmon pink) and $N_{\rm bin}=3$ (brown). For clarity, the plotted range of the error contours is changed in each panel. }
  \label{tomo_HSC_LSST}
\end{figure*}
\begin{table*}[ht]
\begin{ruledtabular}
  \begin{tabular}{|l|l|l|l|l|}
    &HSC~($N_{\rm bin}=1$)&HSC~($N_{\rm bin}=3$)&LSST~($N_{\rm bin}=1$)&LSST~($N_{\rm bin}=3$)\\ \hline
    $\sigma (\fnl)$&$29\ (10)$&$12\ (6.1)$&$2.2(0.89)$&$1.7(0.89)$\\ 
    $\sigma (\gnl)$&$1.4\times 10^5\ (8.0\times 10^4)$&$5.2\times 10^4\ (3.4\times 10^4)$&$5.4\times 10^3\ (3.8\times 10^3)$&$4.4\times 10^3\ (3.3\times 10^3)$\\
   $\sigma (\tnl)$&$6.1\times 10^3\ (2.5\times 10^3)$&$2.0\times 10^3\ (1.0\times 10^3)$&$15\ (6.4)$&$11\ (6.1)$\\ 
 \end{tabular}
  \caption{Forecast results of marginalized (un-marginalized) $1\sigma$ errors on primordial non-Gaussian parameters for HSC and LSST, taking account of the tomography. Here, ignoring the weak-lensing measurement, the constraints derived from the halo clustering is simply estimated, assuming the three redshift bins ($N_{\rm bin}=3$, brown). The results are then compared with those from the single-redshift bin ($N_{\rm bin}=1$, salmon pink). }
  \label{tomomo}
  \end{ruledtabular}
  \end{table*}  

\section{Impact of tomographic measurement \label{sec:discuss}}

So far, we have presented the forecast results for the representative surveys assuming the single-redshift bin without tomography. In this section, we discuss the impact of tomographic technique, and demonstrate how well one can improve the constraints on $f_{\rm NL}$, $g_{\rm NL}$ and $\tau_{\rm NL}$. For illustrative purpose, we only consider the power spectrum and bispectrum of halo clustering, ignoring the photo-$z$ error. The tomography including the cross correlation with weak-lensing observations will be considered elsewhere. 
In the absence of weak-lensing observations, 
summing up independently the Fisher matrices of 
Eq.~(\ref{eq:Fisher_total}) computed at each redshift bin yields a combined constraint on the primordial non-Gaussianity from the tomographic measurements. 
Here, we specifically consider the cases with HSC and LSST, whose survey depths are wide enough to apply tomographic technique.

Fig.~\ref{tomo_HSC_LSST} and Table~\ref{tomomo} summarize
the forecast results. 
We divide the observed redshift distribution of halos into three bins, i.e., $N_{\rm bin}=3$, with equal number of halos in each redshift bin. 
The results show  
that the tomographic measurement improves the constraints on primordial non-Gaussianity, and the marginalized constraints become tighter than the no-tomographic case (i.e., $N_{\rm bin}=1$) by a factor of $1.4-3$. This is simply because the 
fractional contribution of primordial non-Gaussianity to the halo clustering becomes large as increasing the redshifts in both power spectrum and bispectrum. In this respect, the redshift dependence of the clustering amplitude is a key to tightly constrain primordial non-Gaussianity. Thus, 
as long as the shot-noise contribution [Eq.~(\ref{covbis})] can be 
sub-dominant, further increasing the number of redshift bins would improve the constraint.  An interesting question may be what is the tightest constraints achievable with the optimal number of redshift bins for a given survey. To precisely answer this question, however, our analysis with the Limber approximation becomes inadequate for narrow redshift bins, and the full-sky treatment, which 
generally requires multi-dimensional numerical integration along the line-of-sight, needs to be implemented for the analytic calculations of power spectrum and bispectrum. Although we postpone a rigorous analysis with full-sky treatment,  
we conclude here that the tomographic technique is important to further constrain primordial non-Gaussianity, and is rather essential to test single-sourced consistency relation from the imaging surveys.

\section{Summary\label{sec:sum}}

In this paper, we have investigated the statistical power of the 
higher-order statistics and cross correlation statistics to detect or constrain primordial non-Gaussianity from imaging surveys. In particular, we have studied 
how well one can simultaneously constrain multiple parameters that characterizes the local-type primordial non-Gaussianity, i.e., $f_{\rm NL}$, $g_{\rm NL}$, and $\tau_{\rm NL}$ from the halo clustering and weak-lensing observations. Adopting the integrated perturbation theory (iPT) which systematically incorporates both the non-Gaussian mode-coupling of primordial density fluctuations and nonlinear halo biasing into theoretical template, we estimate the signal-to-noise ratios of angular-power spectra and bispectra for halo clustering and weak-lensing observations. The contributions from the primordial non-Gaussianity for 
each parameter of $f_{\rm NL}$, $g_{\rm NL}$, and $\tau_{\rm NL}$ are found to 
have different asymptotic scale-dependence in the angular bispectra, and this 
enables us to break the parameter degeneracy found in the power spectrum.

Based on the Fisher matrix analysis, 
the forecast results for the three representative surveys (HSC, DES, and LSST) 
show that the measurement of halo bispectrum can give tighter constraints on 
non-Gaussian parameters than those from the power spectrum, and the marginalized constraints rather approach to the single-parameter (un-marginalized) constraint. As a result, simultaneous constraints become possible with the measurement of halo bispectrum. Further adding the cross correlation with weak-lensing observations moderately improves the constraints, but with a full combination of both the power spectra and bispectra, idealistically deep and wide survey of LSST can give significantly tight constraints on $f_{\rm NL}$, $g_{\rm NL}$, and $\tau_{\rm NL}$, that surpass even the single-parameter constraints derived 
from the Planck CMB measurement. Also, a simple demonstration with the tomographic technique implies that the constraints from the deep photometric surveys will be further improved by a factor of $1.4-3$ for each non-Gaussian parameter.

The forecast results presented in this paper may be conservative in the sense that we have considered only the large-angular scales of $\ell\leq \ell_{\rm max}=150$. For the halo angular bispectra, increasing the maximum multipole potentially leads to a much tighter constraint on the local-type primordial non-Gaussianity, because the number of possible squeezed triangles increases with $\ell_{\rm max}$. For a quantitative estimation, however, the 
gravity-induced contributions to the bispectrum amplitude and its 
error covariance are rather important, and a more careful study is necessary. Another important aspect to improve the constraint is a tomographic technique combining the cross correlation with weak-lensing observations. Since the 
lensing effect induces non-zero correlation between different redshift bins, the number of possible cross-correlation quantities is increased, and this will help to further constrain primordial non-Gaussianity from the deep photometric surveys. To investigate this, a full-sky analysis without using Limber approximation would be essential. 

Finally, our forecast results are derived based on the prediction with iPT, assuming a prior knowledge of halo bias properties. While the power spectrum prediction with iPT has been tested against the halo clustering in $N$-body simulations, the prediction of bispectrum with non-Gaussian initial contributions has not yet been tested. Although iPT is shown to be consistent with previously known analytic treatment, and thus our results are qualitatively correct, the accuracy check of the prediction is important for future measurements. Further, our treatment of halo bias may be optimistic, and for a proper comparison with observations, we need to incorporate nuisance parameters into the characterization of 
halo bias to reduce the impact of unknown systematics. A further quantitative 
study with simulations is definitely necessary toward a practical application.

\acknowledgments
This work is in part supported by MEXT KAKENHI (15H05888 for SY, 15H05889 for AT, and 15H05890 for TM). SY is also supported from the Grants-in-Aid for Scientific Research from the JSPS (No. 15K17659). IH acknowledges a support from the Hayakawa Sachio Fund of the Astronomical Society of Japan, and TN is grateful for a support from the JSPS Postdoctoral Fellowships for Research Abroad (No.~26-142). 

\newpage

\appendix

\section{Power spectra and Bispectra in the large-scale limit 
\label{sec:largelim}}

In this Appendix, we present the expressions for the power spectra and bispectra in the large-scale limit described in Sec.~\ref{sec:multi}. 
First, tree-level and one-loop contributions to the three-dimensional power spectra given in Eq.~(\ref{Ptreee}) respectively become
\begin{widetext}
\begin{align}
\mathcal{P}_{\rm grav}(k)&\simeq\Gamma_{\rm X}^{(1)}(\bm{k})\Gamma_{\rm Y}^{(1)}(\bm{k})P_{\rm L}(k)+\frac{1}{2}\int\frac{d^3 p}{(2\pi)^3}\Gamma_{\rm X}^{(2)}(\bm{p},-\bm{p})\Gamma_{\rm Y}^{(2)}(\bm{p},-\bm{p})P_{\rm L}(k)P_{\rm L}(p),\label{power3D-Lgra}\\
\mathcal{P}^{\rm 1-loop}_{f_{\rm NL}}(k)&\simeq2\Gamma_{\rm X}^{(1)}(\bm{k})\frac{P_{\rm L}(k)}{M(k)}\int\frac{d^3p}{(2\pi)^3}\Gamma_{\rm Y}^{(2)}(-\bm{p},\bm{p})P_{\rm L}(p)+1{\rm perms}(X \leftrightarrow Y),
\end{align}
where $\mathcal{P}^{\rm 1-loop}_{\rm bis}(k)=f_{\rm NL}\mathcal{P}^{\rm 1-loop}_{f_{\rm NL}}$. On the other hand, the un-decomposable two-loop contribution in Eq.~(\ref{2-looppower}) is reduced to 
\begin{align}
\mathcal{P}^{\rm 2-loop}_{\rm \gnl}(k)&\simeq3\Gamma_{\rm X}^{(1)}(\bm{k})\frac{P_{\rm L}(k)}{M(k)}\int\frac{d^3p_1d^3p_2}{(2\pi)^6}\Gamma_{\rm Y}^{(3)}(-\bm{p}_1,-\bm{p}_2,\bm{p}_1+\bm{p}_2)\nonumber\\
&\times M(p_1)M(p_2)M(|\bm{p}_1+\bm{p}_2|)P_{\Phi}(p_1)P_{\Phi}(p_2)+1{\rm perms}(X \leftrightarrow Y),\\
\mathcal{P}^{\rm 2-loop}_{\rm \tnl}(k)&\simeq\frac{25}{9}\frac{P_{\rm L}(k)}{M(k)}\Biggl[\Gamma_{\rm X}^{(1)}(\bm{k})\int\frac{d^3p_1d^3p_2}{(2\pi)^6}\Gamma_{\rm Y}^{(3)}(-\bm{p}_1,-\bm{p}_2,\bm{p}_1+\bm{p}_2)\nonumber\\
&\times M(p_1)M(p_2)M(|\bm{p}_1+\bm{p}_2|)P_{\Phi}(p_1)P_{\Phi}(p_2)\nonumber\\
&+\frac{1}{2}\frac{1}{M(k)}\int\frac{d^3p_1d^3p_2}{(2\pi)^6}\Gamma_{\rm X}^{(2)}(\bm{p}_1,-\bm{p}_1)\Gamma_{\rm Y}^{(2)}(\bm{p}_2,-\bm{p}_2)P_{\rm L}({p}_1)P_{\rm L}({p}_2)\Biggr]\nonumber\\
&+1{\rm perms}(X \leftrightarrow Y),
\label{power3D-L}
\end{align}
\end{widetext}
where we define $\mathcal{P}^{\rm 2-loop}_{\rm tris}=g_{\rm NL}\mathcal{P}^{\rm 2-loop}_{\rm \gnl}+\tau_{\rm NL}\mathcal{P}^{\rm 2-loop}_{\rm \tnl}$. Note that the multi-point propagators $\Gamma_X^{(n)}$ in these expressions are evaluated with those summarized in Sec.\ref{sec:multi}.

Similarly, the large-scale limit of the three-dimensional bispectra given in 
Eqs.~(\ref{eq:btree}), (\ref{eq:b1loop}), and (\ref{2-loopbi})
becomes
\begin{widetext}  
\begin{align}
\mathcal{B}^{\rm tree}_{f_{\rm NL}}\left(\bm{k}_1,\bm{k}_2,\bm{k}_3\right)=&2\Gamma^{\left(1\right)}_{\rm X}\left(\bm{k}_1\right)\Gamma^{\left(1\right)}_{\rm Y}\left(\bm{k}_2\right)\Gamma^{\left(1\right)}_{\rm Z}\left(\bm{k}_3\right)M\left(k_1\right)M\left(k_2\right)M\left(k_3\right)\,\,
\left[P_{\Phi}\left(k_1\right)P_\Phi\left(k_2\right)+2{\rm perms}(\bm{k}_1\leftrightarrow \bm{k}_2\leftrightarrow \bm{k}_3)\right],\label{Btreebis-L}\\
\mathcal{B}_{\rm grav}^{\rm 1-loop,1}\left(\bm{k}_1,\bm{k}_2,\bm{k}_3\right)\simeq&\int\frac{d^3p}{\left(2\pi\right)^3}P_{\rm L}\left(p\right)^3\Gamma^{\left(2\right)}_{\rm X}\left(\bm{p},-\bm{p}\right)\Gamma^{\left(2\right)}_{\rm Y}\left(\bm{p},-\bm{p}\right)\Gamma^{\left(2\right)}_{\rm Z}\left(\bm{p},-\bm{p}\right),\label{Bl1g-L}\\
\mathcal{B}_{\rm grav}^{\rm 1-loop,2}\left(\bm{k}_1,\bm{k}_2,\bm{k}_3\right)\simeq&\frac{1}{3}\biggl[\biggl\{\Gamma^{\left(1\right)}_{\rm X}\left(\bm{k}_1\right)P_{\rm L}\left(k_1\right)\int\frac{d^3p}{\left(2\pi\right)^3}\Gamma^{\left(2\right)}_{\rm Y}\left(\bm{p},-\bm{p}\right)\Gamma^{\left(3\right)}_{\rm Z}\left(-\bm{k}_1,-\bm{p},-\bm{k}_2+\bm{p}\right)P_{\rm L}\left(p\right)^2\nonumber\\
+&2{\rm perms}({\rm X\leftrightarrow Y\leftrightarrow Z})\biggr\}+5{\rm perms}(\bm{k}_1\leftrightarrow \bm{k}_2\leftrightarrow \bm{k}_3)\biggr],\label{Bl2g-L}\\
\mathcal{B}_{f_{\rm NL}}^{\rm 1-loop,1}\left(\bm{k}_1,\bm{k}_2,\bm{k}_3\right)\simeq&\frac{4}{3}\biggl[\biggl\{\Gamma^{\left(1\right)}_{\rm X}\left(\bm{k}_1\right)\Gamma^{\left(1\right)}_{\rm Y}\left(\bm{k}_2\right)P_{\rm L}\left(k_1\right)\frac{P_{\rm L}\left(k_2\right)}{M\left(k_2\right)}\int\frac{d^3p}{\left(2\pi\right)^3}\Gamma^{\left(3\right)}_{\rm Z}\left(-\bm{k}_1,-\bm{p},-\bm{k}_2+\bm{p}\right)P_{\rm L}\left(p\right)\nonumber\\
+&2{\rm perms}({\rm X\leftrightarrow Y\leftrightarrow Z})\biggr\}+5{\rm perms}(\bm{k}_1\leftrightarrow \bm{k}_2\leftrightarrow \bm{k}_3)\biggr],\label{Bl1b-L}\\
\mathcal{B}_{f_{\rm NL}}^{\rm 1-loop,2}\left(\bm{k}_1,\bm{k}_2,\bm{k}_3\right)\simeq&\frac{4}{3}\biggl[\Gamma^{\left(1\right)}_{\rm X}\left(\bm{k}_1\right)\Gamma^{\left(2\right)}_{\rm Y}\left(\bm{k}_1,\bm{k}_2\right)P_{\rm L}\left(k_1\right)\frac{P_{\rm L}\left(k_2\right)}{M\left(k_2\right)}\int\frac{d^3p}{\left(2\pi\right)^3}\Gamma^{\left(2\right)}_{\rm Z}\left(\bm{p},-\bm{p}\right)P_{\rm L}\left(p\right)\nonumber\\
+&2{\rm perms}({\rm X\leftrightarrow Y\leftrightarrow Z})\biggr\}+5{\rm perms}(\bm{k}_1\leftrightarrow \bm{k}_2\leftrightarrow \bm{k}_3)\biggr],\label{Bl2b-L}\\
\mathcal{B}_{f_{\rm NL}}^{\rm 1-loop,3}\left(\bm{k}_1,\bm{k}_2,\bm{k}_3\right)\simeq&\frac{4}{3}\biggl[\biggl\{\Gamma^{\left(1\right)}_{\rm X}\left(\bm{k}_1\right)\frac{P_{\rm L}\left(k_1\right)}{M\left(k_1\right)}\int\frac{d^3p}{\left(2\pi\right)^3}\Gamma^{\left(2\right)}_{\rm Y}\left(\bm{p},-\bm{p}\right)\Gamma^{\left(2\right)}_{\rm Z}\left(\bm{p},-\bm{p}\right)P_{\rm L}\left(p\right)^2\nonumber\\
+&2{\rm perms}({\rm X\leftrightarrow Y\leftrightarrow Z})\biggr\}+2{\rm perms}(\bm{k}_1\leftrightarrow \bm{k}_2\leftrightarrow \bm{k}_3)\biggr],\label{BBBBB}\\
\mathcal{B}^{\rm 1-loop}_{g_{\rm NL}}\left(\bm{k}_1,\bm{k}_2,\bm{k}_3\right)\simeq&2\Biggl[\Biggl\{\Gamma^{\left(1\right)}_{\rm X}\left(\bm{k}_1\right)\Gamma^{\left(1\right)}_{\rm Y}\left(\bm{k}_2\right)M\left(k_1\right)M\left(k_2\right)P_{\Phi}\left(k_1\right)P_\Phi\left(k_2\right)\int\frac{d^3p}{\left(2\pi\right)^3}\Gamma^{\left(2\right)}_{\rm Z}\left(\bm{p},-\bm{p}\right)P_{\rm L}\left(p\right)\nonumber\\
+&2{\rm perms}({\rm X\leftrightarrow Y\leftrightarrow Z})\Biggr\}+2{\rm perms}(\bm{k}_1\leftrightarrow \bm{k}_2\leftrightarrow \bm{k}_3)\Biggr],\label{1-loopgnlB}\\
\mathcal{B}^{\rm 1-loop}_{\tau_{\rm NL}}\left(\bm{k}_1,\bm{k}_2,\bm{k}_3\right)\simeq&\frac{25}{27}\biggl[\int\frac{d^3p}{\left(2\pi\right)^3}\Gamma^{\left(2\right)}_{\rm Z}\left(\bm{p},-\bm{p}\right)P_{\rm L}\left(p\right)\nonumber\\
\times&\biggl\{\Gamma^{\left(1\right)}_{\rm X}\left(\bm{k}_1\right)\Gamma^{\left(1\right)}_{\rm Y}\left(\bm{k}_2\right)M\left(k_1\right)M\left(k_2\right)\left[P_{\Phi}\left(k_1\right)P_\Phi\left(k_2\right)+P_\Phi\left(k_2\right)P_\Phi\left(k_3\right)+P_{\Phi}\left(k_1\right)P_\Phi\left(k_3\right)\right]\nonumber\\
+&\Gamma^{\left(1\right)}_{\rm X}\left(\bm{k}_1\right)\Gamma^{\left(1\right)}_{\rm Y}\left(\bm{k}_2\right)M\left(k_1\right)M\left(k_2\right)\left[P_{\Phi}\left(k_1\right)+P_{\Phi}\left(k_2\right)\right]P_{\Phi}\left(p\right)\nonumber\\
+&2{\rm perms}(\bm{k}_1\leftrightarrow \bm{k}_2\leftrightarrow \bm{k}_3)\biggr\}+2{\rm perms}({\rm X\leftrightarrow Y\leftrightarrow Z})\biggr],\\
\mathcal{B}^{\rm 2-loop}_{g_{\rm NL}}\left(\bm{k}_1,\bm{k}_2,\bm{k}_3\right)\simeq&3\biggl[\biggl\{\Gamma^{\left(1\right)}_{\rm X}\left(\bm{k}_1\right)\Gamma^{\left(2\right)}_{\rm Y}\left(\bm{p}_1,-\bm{p}_1\right)P_{\rm L}\left(k_1\right)\frac{P_{\rm L}\left(k_3\right)}{M(k_3)}\int\frac{d^3p_1d^3p_2}{(2\pi)^6}P_{\rm L}\left(p_1\right)\nonumber\\
\times&\Gamma^{\left(3\right)}_{\rm Z}\left(\bm{p}_1,\bm{p}_2,-\bm{p}_1-\bm{p}_2\right)M(p_1)M(p_2)M(|\bm{p}_1+\bm{p}_2|)P_\phi(p_1)P_\phi(p_2)\nonumber\\
+&5{\rm perms}(\bm{k}_1\leftrightarrow \bm{k}_2\leftrightarrow \bm{k}_3)\biggr\}+2{\rm perms}({\rm X\leftrightarrow Y\leftrightarrow Z})\biggr],\\
\mathcal{B}^{\rm 2-loop}_{\tau_{\rm NL}}\left(\bm{k}_1,\bm{k}_2,\bm{k}_3\right)\simeq&\frac{25}{9}\biggl[\biggl\{\Gamma^{\left(1\right)}_{\rm X}\left(\bm{k}_1\right)\Gamma^{\left(2\right)}_{\rm Y}\left(\bm{p}_1,-\bm{p}_1\right)P_{\rm L}\left(k_1\right)\frac{P_{\rm L}\left(k_3\right)}{M(k_3)}\int\frac{d^3p_1d^3p_2}{(2\pi)^6}P_{\rm L}\left(p_1\right)\nonumber\\
\times&\Gamma^{\left(3\right)}_{\rm Z}\left(\bm{p}_1,\bm{p}_2,-\bm{p}_1-\bm{p}_2\right)M(p_1)M(p_2)M(|\bm{p}_1+\bm{p}_2|)P_\phi(p_1)P_\phi(p_2)\nonumber\\
+&5{\rm perms}(\bm{k}_1\leftrightarrow \bm{k}_2\leftrightarrow \bm{k}_3)\biggr\}+2{\rm perms}({\rm X\leftrightarrow Y\leftrightarrow Z})\biggr],\label{2-looptaunlB}
\end{align}
\end{widetext}
where we define 
\begin{align}
&\mathcal{B}^{\rm tree}_{\rm bis}=\fnl\mathcal{B}^{\rm tree}_{f_{\rm NL}}, 
\nonumber\\
&\mathcal{B}^{\rm 1-loop,i}_{\rm bis}=\fnl\mathcal{B}^{\rm 1-loop,i}_{f_{\rm NL}}, 
\nonumber\\
&\mathcal{B}^{\rm 1-loop}_{\rm tris}=g_{\rm NL}\mathcal{B}^{\rm 1-loop}_{\rm \gnl}+\tau_{\rm NL}\mathcal{B}^{\rm 1-loop}_{\rm \tnl}
\nonumber\\
&\mathcal{B}^{\rm 2-loop}_{\rm tris}=g_{\rm NL}\mathcal{B}^{\rm 2-loop}_{\rm \gnl}+\tau_{\rm NL}\mathcal{B}^{\rm 2-loop}_{\rm \tnl}.
\end{align}

\bibliographystyle{plain}
\bibliography{bib2}

\begin{thebibliography}{10}%
\makeatletter
\providecommand \@ifxundefined [1]{%
 \ifx #1\undefined \expandafter \@firstoftwo
 \else \expandafter \@secondoftwo
\fi
}%
\providecommand \@ifnum [1]{%
 \ifnum #1\expandafter \@firstoftwo
 \else \expandafter \@secondoftwo
\fi
}%
\providecommand \enquote [1]{``#1''}%
\providecommand \bibnamefont  [1]{#1}%
\providecommand \bibfnamefont [1]{#1}%
\providecommand \citenamefont [1]{#1}%
\providecommand\href[0]{\@sanitize\@href}%
\providecommand\@href[1]{\endgroup\@@startlink{#1}\endgroup\@@href}%
\providecommand\@@href[1]{#1\@@endlink}%
\providecommand \@sanitize [0]{\begingroup\catcode`\&12\catcode`\#12\relax}%
\@ifxundefined \pdfoutput {\@firstoftwo}{%
 \@ifnum{\z@=\pdfoutput}{\@firstoftwo}{\@secondoftwo}%
}{%
 \providecommand\@@startlink[1]{\leavevmode\special{html:<a href="#1">}}%
 \providecommand\@@endlink[0]{\special{html:</a>}}%
}{%
 \providecommand\@@startlink[1]{%
  \leavevmode
  \pdfstartlink
   attr{/Border[0 0 1 ]/H/I/C[0 1 1]}%
   user{/Subtype/Link/A<</Type/Action/S/URI/URI(#1)>>}%
  \relax
 }%
 \providecommand\@@endlink[0]{\pdfendlink}%
}%
\providecommand \url  [0]{\begingroup\@sanitize \@url }%
\providecommand \@url [1]{\endgroup\@href {#1}{\urlprefix}}%
\providecommand \urlprefix [0]{URL }%
\providecommand \Eprint[0]{\href }%
\@ifxundefined \urlstyle {%
  \providecommand \doi [1]{doi:\discretionary{}{}{}#1}%
}{%
  \providecommand \doi [0]{doi:\discretionary{}{}{}\begingroup
  \urlstyle{rm}\Url }%
}%
\providecommand \doibase [0]{http://dx.doi.org/}%
\providecommand \Doi[1]{\href{\doibase#1}}%
\providecommand \bibAnnote [3]{%
  \BibitemShut{#1}%
  \begin{quotation}\noindent
    \textsc{Key:}\ #2\\\textsc{Annotation:}\ #3%
  \end{quotation}%
}%
\providecommand \bibAnnoteFile [2]{%
  \IfFileExists{#2}{\bibAnnote {#1} {#2} {\input{#2}}}{}%
}%
\providecommand \typeout [0]{\immediate \write \m@ne }%
\providecommand \selectlanguage [0]{\@gobble}%
\providecommand \bibinfo [0]{\@secondoftwo}%
\providecommand \bibfield [0]{\@secondoftwo}%
\providecommand \translation [1]{[#1]}%
\providecommand \BibitemOpen[0]{}%
\providecommand \bibitemStop [0]{}%
\providecommand \bibitemNoStop [0]{.\EOS\space}%
\providecommand \EOS [0]{\spacefactor3000\relax}%
\providecommand \BibitemShut [1]{\csname bibitem#1\endcsname}%
\bibitem{2010JCAP...12..030S}%
  \BibitemOpen
  \bibfield{author}{%
  \bibinfo {author} {\bibfnamefont{T.}~\bibnamefont{{Suyama}}}, \bibinfo
  {author} {\bibfnamefont{T.}~\bibnamefont{{Takahashi}}}, \bibinfo {author}
  {\bibfnamefont{M.}~\bibnamefont{{Yamaguchi}}},\ and\ \bibinfo {author}
  {\bibfnamefont{S.}~\bibnamefont{{Yokoyama}}},\ }%
  \bibfield{journal}{%
  \Doi{10.1088/1475-7516/2010/12/030}{\bibinfo {journal} {JCAP}}\ }%
  \textbf{\bibinfo {volume} {12}},\ \bibinfo {eid} {030} (\bibinfo {month}
  {Dec.}\ \bibinfo {year} {2010}),\
  \Eprint{http://arxiv.org/abs/1009.1979}{arXiv:1009.1979 [astro-ph.CO]}%
  \bibAnnoteFile{NoStop}{2010JCAP...12..030S}%
 \bibitem{2013arXiv1303.5084P}%
  \BibitemOpen
  \bibfield{author}{%
  \bibinfo {author} {\bibnamefont{{Planck Collaboration}}}, \bibinfo {author}
  {\bibfnamefont{P.~A.~R.}\ \bibnamefont{{Ade}}}, \bibinfo {author}
  {\bibfnamefont{N.}~\bibnamefont{{Aghanim}}}, \bibinfo {author}
  {\bibfnamefont{C.}~\bibnamefont{{Armitage-Caplan}}}, \bibinfo {author}
  {\bibfnamefont{M.}~\bibnamefont{{Arnaud}}}, \bibinfo {author}
  {\bibfnamefont{M.}~\bibnamefont{{Ashdown}}}, \bibinfo {author}
  {\bibfnamefont{F.}~\bibnamefont{{Atrio-Barandela}}}, \bibinfo {author}
  {\bibfnamefont{J.}~\bibnamefont{{Aumont}}}, \bibinfo {author}
  {\bibfnamefont{C.}~\bibnamefont{{Baccigalupi}}}, \bibinfo {author}
  {\bibfnamefont{A.~J.}\ \bibnamefont{{Banday}}},\ and\ \bibinfo {author}
  {\bibnamefont{et~al.}},\ }%
  \bibfield{journal}{%
  \bibinfo {journal} {ArXiv e-prints}}%
   (\bibinfo {month} {Mar.}\ \bibinfo {year} {2013}),\
  \Eprint{http://arxiv.org/abs/1303.5084}{arXiv:1303.5084 [astro-ph.CO]}%
  \bibAnnoteFile{NoStop}{2013arXiv1303.5084P}%
\bibitem{2008PhRvD..77b3505S}%
  \BibitemOpen
  \bibfield{author}{%
  \bibinfo {author} {\bibfnamefont{T.}~\bibnamefont{{Suyama}}}\ and\ \bibinfo
  {author} {\bibfnamefont{M.}~\bibnamefont{{Yamaguchi}}},\ }%
  \bibfield{journal}{%
  \Doi{10.1103/PhysRevD.77.023505}{\bibinfo {journal} {Phys. Rev.}}\ }%
  \textbf{\bibinfo {volume} {77}},\ \bibinfo {eid} {023505} (\bibinfo {month}
  {Jan.}\ \bibinfo {year} {2008}),\
  \Eprint{http://arxiv.org/abs/0709.2545}{arXiv:0709.2545}%
  \bibAnnoteFile{NoStop}{2008PhRvD..77b3505S}%
\bibitem{2008PhRvD..77l3514D}%
  \BibitemOpen
  \bibfield{author}{%
  \bibinfo {author} {\bibfnamefont{N.}~\bibnamefont{{Dalal}}}, \bibinfo
  {author} {\bibfnamefont{O.}~\bibnamefont{{Dor{\'e}}}}, \bibinfo {author}
  {\bibfnamefont{D.}~\bibnamefont{{Huterer}}},\ and\ \bibinfo {author}
  {\bibfnamefont{A.}~\bibnamefont{{Shirokov}}},\ }%
  \bibfield{journal}{%
  \Doi{10.1103/PhysRevD.77.123514}{\bibinfo {journal} {Phys. Rev.}}\ }%
  \textbf{\bibinfo {volume} {77}},\ \bibinfo {eid} {123514} (\bibinfo {month}
  {Jun.}\ \bibinfo {year} {2008}),\
  \Eprint{http://arxiv.org/abs/0710.4560}{arXiv:0710.4560}%
  \bibAnnoteFile{NoStop}{2008PhRvD..77l3514D}%
\bibitem{2008PhRvD..78l3507A}%
  \BibitemOpen
  \bibfield{author}{%
  \bibinfo {author} {\bibfnamefont{N.}~\bibnamefont{{Afshordi}}}\ and\ \bibinfo
  {author} {\bibfnamefont{A.~J.}\ \bibnamefont{{Tolley}}},\ }%
  \bibfield{journal}{%
  \Doi{10.1103/PhysRevD.78.123507}{\bibinfo {journal} {Phys. Rev.}}\ }%
  \textbf{\bibinfo {volume} {78}},\ \bibinfo {eid} {123507} (\bibinfo {month}
  {Dec.}\ \bibinfo {year} {2008}),\
  \Eprint{http://arxiv.org/abs/0806.1046}{arXiv:0806.1046}%
  \bibAnnoteFile{NoStop}{2008PhRvD..78l3507A}%
\bibitem{2011PhRvD..83l3514N}%
  \BibitemOpen
  \bibfield{author}{%
  \bibinfo {author} {\bibfnamefont{T.}~\bibnamefont{{Namikawa}}}, \bibinfo
  {author} {\bibfnamefont{T.}~\bibnamefont{{Okamura}}},\ and\ \bibinfo {author}
  {\bibfnamefont{A.}~\bibnamefont{{Taruya}}},\ }%
  \bibfield{journal}{%
  \Doi{10.1103/PhysRevD.83.123514}{\bibinfo {journal} {Phys. Rev.}}\ }%
  \textbf{\bibinfo {volume} {83}},\ \bibinfo {eid} {123514} (\bibinfo {month}
  {Jun.}\ \bibinfo {year} {2011}),\
  \Eprint{http://arxiv.org/abs/1103.1118}{arXiv:1103.1118 [astro-ph.CO]}%
  \bibAnnoteFile{NoStop}{2011PhRvD..83l3514N}%
\bibitem{2012PhRvD..85d3518T}%
  \BibitemOpen
  \bibfield{author}{%
  \bibinfo {author} {\bibfnamefont{Y.}~\bibnamefont{{Takeuchi}}}, \bibinfo
  {author} {\bibfnamefont{K.}~\bibnamefont{{Ichiki}}},\ and\ \bibinfo {author}
  {\bibfnamefont{T.}~\bibnamefont{{Matsubara}}},\ }%
  \bibfield{journal}{%
  \Doi{10.1103/PhysRevD.85.043518}{\bibinfo {journal} {Phys. Rev.}}\ }%
  \textbf{\bibinfo {volume} {85}},\ \bibinfo {eid} {043518} (\bibinfo {month}
  {Feb.}\ \bibinfo {year} {2012}),\
  \Eprint{http://arxiv.org/abs/1111.6835}{arXiv:1111.6835 [astro-ph.CO]}%
  \bibAnnoteFile{NoStop}{2012PhRvD..85d3518T}%
\bibitem{Seljak:2008xr}%
  \BibitemOpen
  \bibfield{author}{%
  \bibinfo {author} {\bibfnamefont{U.}~\bibnamefont{Seljak}},\ }%
  \bibfield{journal}{%
  \Doi{10.1103/PhysRevLett.102.021302}{\bibinfo {journal} {Phys. Rev. Lett.}}\
  }%
  \textbf{\bibinfo {volume} {102}},\ \bibinfo {pages} {021302} (\bibinfo {year}
  {2009}),\ \Eprint{http://arxiv.org/abs/0807.1770}{arXiv:0807.1770
  [astro-ph]}%
  \bibAnnoteFile{NoStop}{Seljak:2008xr}%
\bibitem{2011PhRvD..84h3509H}%
  \BibitemOpen
  \bibfield{author}{%
  \bibinfo {author} {\bibfnamefont{N.}~\bibnamefont{{Hamaus}}}, \bibinfo
  {author} {\bibfnamefont{U.}~\bibnamefont{{Seljak}}},\ and\ \bibinfo {author}
  {\bibfnamefont{V.}~\bibnamefont{{Desjacques}}},\ }%
  \bibfield{journal}{%
  \Doi{10.1103/PhysRevD.84.083509}{\bibinfo {journal} {Phys. Rev.}}\ }%
  \textbf{\bibinfo {volume} {84}},\ \bibinfo {eid} {083509} (\bibinfo {month}
  {Oct.}\ \bibinfo {year} {2011}),\
  \Eprint{http://arxiv.org/abs/1104.2321}{arXiv:1104.2321 [astro-ph.CO]}%
  \bibAnnoteFile{NoStop}{2011PhRvD..84h3509H}%
\bibitem{Yamauchi:2015mja}%
  \BibitemOpen
  \bibfield{author}{%
  \bibinfo {author} {\bibfnamefont{D.}~\bibnamefont{Yamauchi}}\ and\ \bibinfo
  {author} {\bibfnamefont{K.}~\bibnamefont{Takahashi}}}%
   (\bibinfo {year} {2015}),\
  \Eprint{http://arxiv.org/abs/1509.07585}{arXiv:1509.07585 [astro-ph.CO]}%
  \bibAnnoteFile{NoStop}{Yamauchi:2015mja}%
\bibitem{2014PhRvD..89d3524Y}%
  \BibitemOpen
  \bibfield{author}{%
  \bibinfo {author} {\bibfnamefont{S.}~\bibnamefont{{Yokoyama}}}, \bibinfo
  {author} {\bibfnamefont{T.}~\bibnamefont{{Matsubara}}},\ and\ \bibinfo
  {author} {\bibfnamefont{A.}~\bibnamefont{{Taruya}}},\ }%
  \bibfield{journal}{%
  \Doi{10.1103/PhysRevD.89.043524}{\bibinfo {journal} {Phys. Rev.}}\ }%
  \textbf{\bibinfo {volume} {89}},\ \bibinfo {eid} {043524} (\bibinfo {month}
  {Feb.}\ \bibinfo {year} {2014}),\
  \Eprint{http://arxiv.org/abs/1310.4925}{arXiv:1310.4925 [astro-ph.CO]}%
  \bibAnnoteFile{NoStop}{2014PhRvD..89d3524Y}%
\bibitem{2009ApJ...703.1230J}%
  \BibitemOpen
  \bibfield{author}{%
  \bibinfo {author} {\bibfnamefont{D.}\ \bibnamefont{{Jeong}}}, \bibinfo
  {author} {\bibfnamefont{E.}\ \bibnamefont{{Komatsu}}},\ }%
  \bibfield{journal}{%
  \Doi{10.1088/0004-637X/703/2/1230}{\bibinfo {journal} {Astrophys. J.}}\ }%
  \textbf{\bibinfo {volume} {703}},\ \bibinfo {pages} {1230-1248} (\bibinfo {month}
  {Oct.}\ \bibinfo {year} {2009}),\
  \Eprint{http://arxiv.org/abs/0904.0497}{arXiv:0904.0497}%
  \bibAnnoteFile{NoStop}{2009ApJ...703.1230J}%
\bibitem{2010JCAP...07..002N}%
  \BibitemOpen
  \bibfield{author}{%
  \bibinfo {author} {\bibfnamefont{T.}\ \bibnamefont{{Nishimichi}}}, \bibinfo
  {author} {\bibfnamefont{T.}\ \bibnamefont{{Taruya}}}, \bibinfo
  {author} {\bibfnamefont{K.}\ \bibnamefont{{Koyama}}}, \bibinfo
  {author} {\bibfnamefont{C.}\ \bibnamefont{{Sabiu}}} \ }%
  \bibfield{journal}{%
  \Doi{10.1088/1475-7516/2010/07/002}{\bibinfo {journal} {JCAP}}\ }%
  \textbf{\bibinfo {volume} {7}},\ \bibinfo {pages} {002} (\bibinfo {month}
  {Jul.}\ \bibinfo {year} {2010}),\
  \Eprint{http://arxiv.org/abs/0911.4768}{arXiv:0911.4768}%
  \bibAnnoteFile{NoStop}{2010JCAP...07..002N}%
\bibitem{2011JCAP...04..006B}%
  \BibitemOpen
  \bibfield{author}{%
  \bibinfo {author} {\bibfnamefont{T.}\ \bibnamefont{{Baldauf}}}, \bibinfo
  {author} {\bibfnamefont{U.}\ \bibnamefont{{Seljak}}}, \bibinfo
  {author} {\bibfnamefont{L.}\ \bibnamefont{{Senatore}}} \ }%
  \bibfield{journal}{%
  \Doi{10.1088/1475-7516/2011/04/006}{\bibinfo {journal} {JCAP}}\ }%
  \textbf{\bibinfo {volume} {4}},\ \bibinfo {pages} {006} (\bibinfo {month}
  {Apr.}\ \bibinfo {year} {2011}),\
  \Eprint{http://arxiv.org/abs/1011.1513}{arXiv:1011.1513}%
  \bibAnnoteFile{NoStop}{2011JCAP...04..006B}%
\bibitem{2004PhRvD..69j3513S}%
  \BibitemOpen
  \bibfield{author}{%
  \bibinfo {author} {\bibfnamefont{R.}\ \bibnamefont{{Scoccimarro}}}, \bibinfo {author}
  {\bibfnamefont{E.}\ \bibnamefont{{Sefusatti}}}, \bibinfo {author}
  {\bibfnamefont{M.}~\bibnamefont{{Zaldarriaga}}},\ }%
  \bibfield{journal}{%
  \Doi{10.1103/PhysRevD.69.103513}{\bibinfo {journal} {Phys. Rev.}}\ }%
  \textbf{\bibinfo {volume} {69}},\ \bibinfo {pages} {103513} (\bibinfo {month}
  {May}\ \bibinfo {year} {2004}),\
  \Eprint{http://arxiv.org/abs/astro-ph/0312286}{astro-ph/0312286}%
  \bibAnnoteFile{NoStop}{2004PhRvD..69j3513S}%
\bibitem{2007PhRvD..76h3004S}%
  \BibitemOpen
  \bibfield{author}{%
  \bibinfo {author} {\bibfnamefont{E.}\ \bibnamefont{{Sefusatti}}}, \bibinfo
  {author} {\bibfnamefont{E.}\ \bibnamefont{{Komatsu}}},\ }%
  \bibfield{journal}{%
  \Doi{10.1103/PhysRevD.76.083004}{\bibinfo {journal} {Phys. Rev.}}\ }%
  \textbf{\bibinfo {volume} {76}},\ \bibinfo {pages} {083004} (\bibinfo {month}
  {Oct.}\ \bibinfo {year} {2007}),\
  \Eprint{http://arxiv.org/abs/0705.0343}{arXiv:0705.0343}%
  \bibAnnoteFile{NoStop}{2007PhRvD..76h3004S}%
\bibitem{2012MNRAS.425.2903S}%
  \BibitemOpen
  \bibfield{author}{%
  \bibinfo {author} {\bibfnamefont{E.}\ \bibnamefont{{Sefusatti}}}, \bibinfo
  {author} {\bibfnamefont{M.}\ \bibnamefont{{Crocce}}}, \bibinfo
  {author} {\bibfnamefont{V.}\ \bibnamefont{{Desjacques}}},\ }%
  \bibfield{journal}{%
  \Doi{10.1111/j.1365-2966.2012.21271.x}{\bibinfo {journal} {Mon. Not. Roy. Astron. Soc.}}\ }%
  \textbf{\bibinfo {volume} {425}},\ \bibinfo {pages} {2903-2930} (\bibinfo {month}
  {Oct.}\ \bibinfo {year} {2012}),\
  \Eprint{http://arxiv.org/abs/1111.6966}{arXiv:1111.6966}%
  \bibAnnoteFile{NoStop}{2012MNRAS.425.2903S}%
\bibitem{2011PhRvD..83h3518M}%
  \BibitemOpen
  \bibfield{author}{%
  \bibinfo {author} {\bibfnamefont{T.}~\bibnamefont{{Matsubara}}},\ }%
  \bibfield{journal}{%
  \Doi{10.1103/PhysRevD.83.083518}{\bibinfo {journal} {Phys. Rev.}}\ }%
  \textbf{\bibinfo {volume} {83}},\ \bibinfo {eid} {083518} (\bibinfo {month}
  {Apr.}\ \bibinfo {year} {2011}),\
  \Eprint{http://arxiv.org/abs/1102.4619}{arXiv:1102.4619 [astro-ph.CO]}%
  \bibAnnoteFile{NoStop}{2011PhRvD..83h3518M}%
\bibitem{2012PhRvD..86f3518M}%
  \BibitemOpen
  \bibfield{author}{%
  \bibinfo {author} {\bibfnamefont{T.}~\bibnamefont{{Matsubara}}},\ }%
  \bibfield{journal}{%
  \Doi{10.1103/PhysRevD.86.063518}{\bibinfo {journal} {Phys. Rev.}}\ }%
  \textbf{\bibinfo {volume} {86}},\ \bibinfo {eid} {063518} (\bibinfo {month}
  {Sep.}\ \bibinfo {year} {2012}),\
  \Eprint{http://arxiv.org/abs/1206.0562}{arXiv:1206.0562 [astro-ph.CO]}%
  \bibAnnoteFile{NoStop}{2012PhRvD..86f3518M}%
\bibitem{2008PhRvD..78j3521B}%
  \BibitemOpen
  \bibfield{author}{%
  \bibinfo {author} {\bibfnamefont{F.}~\bibnamefont{{Bernardeau}}}, \bibinfo
  {author} {\bibfnamefont{M.}~\bibnamefont{{Crocce}}},\ and\ \bibinfo {author}
  {\bibfnamefont{R.}~\bibnamefont{{Scoccimarro}}},\ }%
  \bibfield{journal}{%
  \Doi{10.1103/PhysRevD.78.103521}{\bibinfo {journal} {Phys. Rev.}}\ }%
  \textbf{\bibinfo {volume} {78}},\ \bibinfo {eid} {103521} (\bibinfo {month}
  {Nov.}\ \bibinfo {year} {2008}),\
  \Eprint{http://arxiv.org/abs/0806.2334}{arXiv:0806.2334}%
  \bibAnnoteFile{NoStop}{2008PhRvD..78j3521B}%
\bibitem{2000ApJ...538..473L}%
  \BibitemOpen
  \bibfield{author}{%
  \bibinfo {author} {\bibfnamefont{A.}~\bibnamefont{{Lewis}}}, \bibinfo
  {author} {\bibfnamefont{A.}~\bibnamefont{{Challinor}}},\ and\ \bibinfo
  {author} {\bibfnamefont{A.}~\bibnamefont{{Lasenby}}},\ }%
  \bibfield{journal}{%
  \Doi{10.1086/309179}{\bibinfo {journal} {Astrophys. J.}}\ }%
  \textbf{\bibinfo {volume} {538}},\ \bibinfo {pages} {473} (\bibinfo {month}
  {Aug.}\ \bibinfo {year} {2000}),\
  \Eprint{http://arxiv.org/abs/astro-ph/9911177}{astro-ph/9911177}%
  \bibAnnoteFile{NoStop}{2000ApJ...538..473L}%
\bibitem{2013ApJS..208...19H}%
  \BibitemOpen
  \bibfield{author}{%
  \bibinfo {author} {\bibfnamefont{G.}~\bibnamefont{{Hinshaw}}}, \bibinfo
  {author} {\bibfnamefont{D.}~\bibnamefont{{Larson}}}, \bibinfo {author}
  {\bibfnamefont{E.}~\bibnamefont{{Komatsu}}}, \bibinfo {author}
  {\bibfnamefont{D.~N.}\ \bibnamefont{{Spergel}}}, \bibinfo {author}
  {\bibfnamefont{C.~L.}\ \bibnamefont{{Bennett}}}, \bibinfo {author}
  {\bibfnamefont{J.}~\bibnamefont{{Dunkley}}}, \bibinfo {author}
  {\bibfnamefont{M.~R.}\ \bibnamefont{{Nolta}}}, \bibinfo {author}
  {\bibfnamefont{M.}~\bibnamefont{{Halpern}}}, \bibinfo {author}
  {\bibfnamefont{R.~S.}\ \bibnamefont{{Hill}}}, \bibinfo {author}
  {\bibfnamefont{N.}~\bibnamefont{{Odegard}}}, \bibinfo {author}
  {\bibfnamefont{L.}~\bibnamefont{{Page}}}, \bibinfo {author}
  {\bibfnamefont{K.~M.}\ \bibnamefont{{Smith}}}, \bibinfo {author}
  {\bibfnamefont{J.~L.}\ \bibnamefont{{Weiland}}}, \bibinfo {author}
  {\bibfnamefont{B.}~\bibnamefont{{Gold}}}, \bibinfo {author}
  {\bibfnamefont{N.}~\bibnamefont{{Jarosik}}}, \bibinfo {author}
  {\bibfnamefont{A.}~\bibnamefont{{Kogut}}}, \bibinfo {author}
  {\bibfnamefont{M.}~\bibnamefont{{Limon}}}, \bibinfo {author}
  {\bibfnamefont{S.~S.}\ \bibnamefont{{Meyer}}}, \bibinfo {author}
  {\bibfnamefont{G.~S.}\ \bibnamefont{{Tucker}}}, \bibinfo {author}
  {\bibfnamefont{E.}~\bibnamefont{{Wollack}}},\ and\ \bibinfo {author}
  {\bibfnamefont{E.~L.}\ \bibnamefont{{Wright}}},\ }%
  \bibfield{journal}{%
  \Doi{10.1088/0067-0049/208/2/19}{\bibinfo {journal} {Astrophys. J. Suppl.}}\
  }%
  \textbf{\bibinfo {volume} {208}},\ \bibinfo {eid} {19} (\bibinfo {month}
  {Oct.}\ \bibinfo {year} {2013}),\
  \Eprint{http://arxiv.org/abs/1212.5226}{arXiv:1212.5226 [astro-ph.CO]}%
  \bibAnnoteFile{NoStop}{2013ApJS..208...19H}%
\bibitem{2010CQGra..27l4002W}%
  \BibitemOpen
  \bibfield{author}{%
  \bibinfo {author} {\bibfnamefont{D.}~\bibnamefont{{Wands}}},\ }%
  \bibfield{journal}{%
  \Doi{10.1088/0264-9381/27/12/124002}{\bibinfo {journal} {Classical and
  Quantum Gravity}}\ }%
  \textbf{\bibinfo {volume} {27}},\ \bibinfo {eid} {124002} (\bibinfo {month}
  {Jun.}\ \bibinfo {year} {2010}),\
  \Eprint{http://arxiv.org/abs/1004.0818}{arXiv:1004.0818}%
  \bibAnnoteFile{NoStop}{2010CQGra..27l4002W}%
\bibitem{2010CQGra..27l4001K}%
  \BibitemOpen
  \bibfield{author}{%
  \bibinfo {author} {\bibfnamefont{K.}~\bibnamefont{{Koyama}}},\ }%
  \bibfield{journal}{%
  \Doi{10.1088/0264-9381/27/12/124001}{\bibinfo {journal} {Classical and
  Quantum Gravity}}\ }%
  \textbf{\bibinfo {volume} {27}},\ \bibinfo {eid} {124001} (\bibinfo {month}
  {Jun.}\ \bibinfo {year} {2010}),\
  \Eprint{http://arxiv.org/abs/1002.0600}{arXiv:1002.0600 [hep-th]}%
  \bibAnnoteFile{NoStop}{2010CQGra..27l4001K}%
\bibitem{2013PhRvD..87b3525Y}%
  \BibitemOpen
  \bibfield{author}{%
  \bibinfo {author} {\bibfnamefont{S.}~\bibnamefont{{Yokoyama}}}\ and\ \bibinfo
  {author} {\bibfnamefont{T.}~\bibnamefont{{Matsubara}}},\ }%
  \bibfield{journal}{%
  \Doi{10.1103/PhysRevD.87.023525}{\bibinfo {journal} {Phys. Rev.}}\ }%
  \textbf{\bibinfo {volume} {87}},\ \bibinfo {eid} {023525} (\bibinfo {month}
  {Jan.}\ \bibinfo {year} {2013}),\
  \Eprint{http://arxiv.org/abs/1210.2495}{arXiv:1210.2495 [astro-ph.CO]}%
  \bibAnnoteFile{NoStop}{2013PhRvD..87b3525Y}%
\bibitem{2001MNRAS.323....1S}%
  \BibitemOpen
  \bibfield{author}{%
  \bibinfo {author} {\bibfnamefont{R.~K.}\ \bibnamefont{{Sheth}}}, \bibinfo
  {author} {\bibfnamefont{H.~J.}\ \bibnamefont{{Mo}}},\ and\ \bibinfo {author}
  {\bibfnamefont{G.}~\bibnamefont{{Tormen}}},\ }%
  \bibfield{journal}{%
  \Doi{10.1046/j.1365-8711.2001.04006.x}{\bibinfo {journal} {Mon. Not. Roy.
  Astron. Soc.}}\ }%
  \textbf{\bibinfo {volume} {323}},\ \bibinfo {pages} {1} (\bibinfo {month}
  {May}\ \bibinfo {year} {2001}),\
  \Eprint{http://arxiv.org/abs/astro-ph/9907024}{astro-ph/9907024}%
  \bibAnnoteFile{NoStop}{2001MNRAS.323....1S}%
\bibitem{HSCrev}%
  \BibitemOpen
  \bibfield{author}{%
  \bibinfo {author} {\bibfnamefont{H.}~\bibnamefont{Collaboration}}}%
   (\bibinfo {month} {May}\ \bibinfo {year} {2009})%
  \bibAnnoteFile{NoStop}{HSCrev}%
\bibitem{2005astro.ph.10346T}%
  \BibitemOpen
  \bibfield{author}{%
  \bibinfo {author} {\bibnamefont{{The Dark Energy Survey Collaboration}}},\ }%
  \bibfield{journal}{%
  \bibinfo {journal} {ArXiv Astrophysics e-prints}}%
   (\bibinfo {month} {Oct.}\ \bibinfo {year} {2005}),\
  \Eprint{http://arxiv.org/abs/astro-ph/0510346}{astro-ph/0510346}%
  \bibAnnoteFile{NoStop}{2005astro.ph.10346T}%
\bibitem{2009arXiv0912.0201L}%
  \BibitemOpen
  \bibfield{author}{%
  \bibinfo {author} {\bibnamefont{{LSST Science Collaboration}}}, \bibinfo
  {author} {\bibfnamefont{P.~A.}\ \bibnamefont{{Abell}}}, \bibinfo {author}
  {\bibfnamefont{J.}~\bibnamefont{{Allison}}}, \bibinfo {author}
  {\bibfnamefont{S.~F.}\ \bibnamefont{{Anderson}}}, \bibinfo {author}
  {\bibfnamefont{J.~R.}\ \bibnamefont{{Andrew}}}, \bibinfo {author}
  {\bibfnamefont{J.~R.~P.}\ \bibnamefont{{Angel}}}, \bibinfo {author}
  {\bibfnamefont{L.}~\bibnamefont{{Armus}}}, \bibinfo {author}
  {\bibfnamefont{D.}~\bibnamefont{{Arnett}}}, \bibinfo {author}
  {\bibfnamefont{S.~J.}\ \bibnamefont{{Asztalos}}}, \bibinfo {author}
  {\bibfnamefont{T.~S.}\ \bibnamefont{{Axelrod}}},\ and\ \bibinfo {author}
  {\bibnamefont{et~al.}},\ }%
  \bibfield{journal}{%
  \bibinfo {journal} {ArXiv e-prints}}%
   (\bibinfo {month} {Dec.}\ \bibinfo {year} {2009}),\
  \Eprint{http://arxiv.org/abs/0912.0201}{arXiv:0912.0201 [astro-ph.IM]}%
  \bibAnnoteFile{NoStop}{2009arXiv0912.0201L}%
\bibitem{1954ApJ...119..655L}%
  \BibitemOpen
  \bibfield{author}{%
  \bibinfo {author} {\bibfnamefont{D.~N.}\ \bibnamefont{{Limber}}},\ }%
  \bibfield{journal}{%
  \Doi{10.1086/145870}{\bibinfo {journal} {Astrophys. J.}}\ }%
  \textbf{\bibinfo {volume} {119}},\ \bibinfo {pages} {655} (\bibinfo {month}
  {May}\ \bibinfo {year} {1954})%
  \bibAnnoteFile{NoStop}{1954ApJ...119..655L}%
\bibitem{2008PhRvD..78l3506L}%
  \BibitemOpen
  \bibfield{author}{%
  \bibinfo {author} {\bibfnamefont{M.}~\bibnamefont{{Loverde}}}\ and\ \bibinfo
  {author} {\bibfnamefont{N.}~\bibnamefont{{Afshordi}}},\ }%
  \bibfield{journal}{%
  \Doi{10.1103/PhysRevD.78.123506}{\bibinfo {journal} {Phys. Rev.}}\ }%
  \textbf{\bibinfo {volume} {78}},\ \bibinfo {eid} {123506} (\bibinfo {month}
  {Dec.}\ \bibinfo {year} {2008}),\
  \Eprint{http://arxiv.org/abs/0809.5112}{arXiv:0809.5112}%
  \bibAnnoteFile{NoStop}{2008PhRvD..78l3506L}%
\bibitem{2013MNRAS.429..344K}%
  \BibitemOpen
  \bibfield{author}{%
  \bibinfo {author} {\bibfnamefont{I.}~\bibnamefont{{Kayo}}}, \bibinfo {author}
  {\bibfnamefont{M.}~\bibnamefont{{Takada}}},\ and\ \bibinfo {author}
  {\bibfnamefont{B.}~\bibnamefont{{Jain}}},\ }%
  \bibfield{journal}{%
  \Doi{10.1093/mnras/sts340}{\bibinfo {journal} {Mon. Not. Roy. Astron. Soc.}}\
  }%
  \textbf{\bibinfo {volume} {429}},\ \bibinfo {pages} {344} (\bibinfo {month}
  {Feb.}\ \bibinfo {year} {2013}),\
  \Eprint{http://arxiv.org/abs/1207.6322}{arXiv:1207.6322 [astro-ph.CO]}%
  \bibAnnoteFile{NoStop}{2013MNRAS.429..344K}%
\bibitem{2013PhR...530...87W}%
  \BibitemOpen
  \bibfield{author}{%
  \bibinfo {author} {\bibfnamefont{D.~H.}\ \bibnamefont{{Weinberg}}}, \bibinfo
  {author} {\bibfnamefont{M.~J.}\ \bibnamefont{{Mortonson}}}, \bibinfo {author}
  {\bibfnamefont{D.~J.}\ \bibnamefont{{Eisenstein}}}, \bibinfo {author}
  {\bibfnamefont{C.}~\bibnamefont{{Hirata}}}, \bibinfo {author}
  {\bibfnamefont{A.~G.}\ \bibnamefont{{Riess}}},\ and\ \bibinfo {author}
  {\bibfnamefont{E.}~\bibnamefont{{Rozo}}},\ }%
  \bibfield{journal}{%
  \Doi{10.1016/j.physrep.2013.05.001}{\bibinfo {journal} {Phys.rep}}\ }%
  \textbf{\bibinfo {volume} {530}},\ \bibinfo {pages} {87} (\bibinfo {month}
  {Sep.}\ \bibinfo {year} {2013}),\
  \Eprint{http://arxiv.org/abs/1201.2434}{arXiv:1201.2434}%
  \bibAnnoteFile{NoStop}{2013PhR...530...87W}%
\bibitem{Yokoyama:2011qr}%
  \BibitemOpen
  \bibfield{author}{%
  \bibinfo {author} {\bibfnamefont{S.}~\bibnamefont{Yokoyama}},\ }%
  \bibfield{journal}{%
  \Doi{10.1088/1475-7516/2011/11/001}{\bibinfo {journal} {JCAP}}\ }%
  \textbf{\bibinfo {volume} {11}},\ \bibinfo {pages} {001} (\bibinfo {year}
  {2011}),\ \Eprint{http://arxiv.org/abs/1108.5569}{arXiv:1108.5569
  [astro-ph.CO]}%
  \bibAnnoteFile{NoStop}{Yokoyama:2011qr}%
\end{thebibliography}%

\end{document}